\documentclass[12pt]{iopart}
\usepackage[dvips]{graphicx}
\usepackage{rotating}
\usepackage{braket}
\usepackage{svg}
\usepackage{soul}
\usepackage{float}
\usepackage{qcircuit}
\usepackage{hyperref}
\usepackage{booktabs}
\usepackage[section]{placeins}
\hypersetup{colorlinks = false, citecolor = black, urlcolor = black}

\begin{document}
\newcommand{\beginsupplement}{%
        \setcounter{table}{0}
        \renewcommand{\thetable}{S\arabic{table}}%
        \setcounter{figure}{0}
        \renewcommand{\thefigure}{S\arabic{figure}}%
}
% \DeclareRobustCommand{\sst}[1]{\texorpdfstring{\st{#1}}{#1}}
% \DeclarePairedDelimiter{\ceil}{\lceil}{\rceil}

\title[Quantum annealing and protein lattice folding]{Investigating the potential for a limited quantum speedup on protein lattice problems}

\author{Carlos Outeiral$^{1,  2}$, Garrett M. Morris$^1$, Jiye Shi$^3$, Martin Strahm$^4$, Simon C. Benjamin$^{2, *}$, Charlotte M. Deane$^{1, \dagger}$}

\address{$^1$Department of Statistics, University of Oxford, 24-29 St Giles', Oxford OX1 3LB, United Kingdom}
\address{$^2$Department of Materials, University of Oxford, Parks  Road,  Oxford OX1 3PH,  United  Kingdom}
\address{$^3$Computer-Aided Drug Design, UCB Pharma, 216 Bath Road, Slough SL1 3WE, United Kingdom}
\address{$^4$Pharma Research \& Early Development, F. Hoffmann-La Roche, Grenzacherstrasse 4058, Basel, Switzerland}
\address{$^*$To whom correspondence shall be addressed: simon.benjamin@materials.ox.ac.uk}
\address{$^\dagger$To whom correspondence shall be addressed: deane@stats.ox.ac.uk}

\begin{abstract}
Protein folding is a central challenge in computational biology, with important applications in molecular biology, drug discovery and catalyst design. As a hard combinatorial optimisation problem, it has been studied as a potential target problem for quantum annealing. Although several experimental implementations have been discussed in the literature, the computational scaling of these approaches has not been elucidated. In this article, we present a numerical study of quantum annealing applied to a large number of small peptide folding problems, aiming to infer useful insights for near-term applications.  We present two conclusions: that even na\"ive quantum annealing, when applied to protein lattice folding, has the potential to outperform classical approaches, and that careful engineering of the Hamiltonians and schedules involved can deliver notable relative improvements for this problem. Overall, our results suggest that quantum algorithms may well offer improvements for problems in the protein folding and structure prediction realm.
\end{abstract}

\vspace{2pc}
\noindent{\it Keywords}: quantum annealing, protein folding, quantum optimisation, biophysics

\section{Introduction}

The structure of a protein captures crucial information about its biological function and therapeutic potential \cite{nelson2008lehninger}. Knowledge of a proteins' structure unlocks valuable biological information, ranging from the ability to predict protein-protein interactions \cite{jones1996principles}, to structure-based discovery of new drugs \cite{jhoti2007structure} and catalysts \cite{zeymer2018directed}. Unfortunately, experiments to determine protein structure are challenging and require extensive time and resources \cite{nelson2008lehninger,blundell2017protein,renaud2018cryo}. As of May 2021, the TrEMBL database \cite{uniprot2019uniprot} contained 214 million protein sequences, while only 177,000 protein structures have been deposited in the Protein Data Bank \cite{berman2003protein}. A reliable computational algorithm for the template-free { -- that is, when a structurally similar protein, perhaps an evolutionary-related specimen from a different organism, is not available to use as a reference \cite{exploringbioinformatics} --} prediction of a protein's structure and its folding pathway from sequence information alone would enable annotation of millions of proteins and could stimulate major advances in biological research. However, despite steady improvements in the past six decades \cite{kendrew1958three,dill2012protein,abriata2019further}, and recent advances in the past year \cite{alphafold2}, a consistent and accurate algorithm for protein folding  from sequence has remained elusive.

Over the past decade, there have been attempts to leverage quantum computing for protein structure prediction \cite{outeiral2020prospects}. The biological structure of a protein is thought to correspond to the minimum of a free energy hypersurface, which for even small peptides is too vast for any classical computer to explore exhaustively \cite{dill2012protein}. A type of quantum computation that may be appropriate to help is quantum annealing, an approach to exploiting the physics of a controlled quantum system that is considered to be of potential use in optimisation problems (whether classical or quantum in nature) \cite{farhi2000quantum}. Typically, the set of possible solutions is mapped to a register of qubits with a binary encoding, and the objective function is represented as a physical Hamiltonian, $H_\mathrm{problem}$, whose eigenvectors and eigenvalues are respectively problem solutions and their scores. In particular, the ground state $\ket{\Phi}$ (or the respective ground eigenspace) corresponds to the global minimum of the problem. The algorithm starts by initialising the register of qubits in the ground state $\ket{\Psi(0)}$ of a given Hamiltonian, $H_\mathrm{trivial}$, whose ground state is easy to prepare, and gently transforming into the problem Hamiltonian, $H_\mathrm{problem}$. If the evolution is slow enough, the adiabatic theorem of quantum mechanics \cite{born1928beweis} ensures that the final state $\ket{\Psi(T)}$ will be infinitesimally close to $\ket{\Phi}$.

\begin{figure}[!t]
    \centering
    \includegraphics[width=\linewidth]{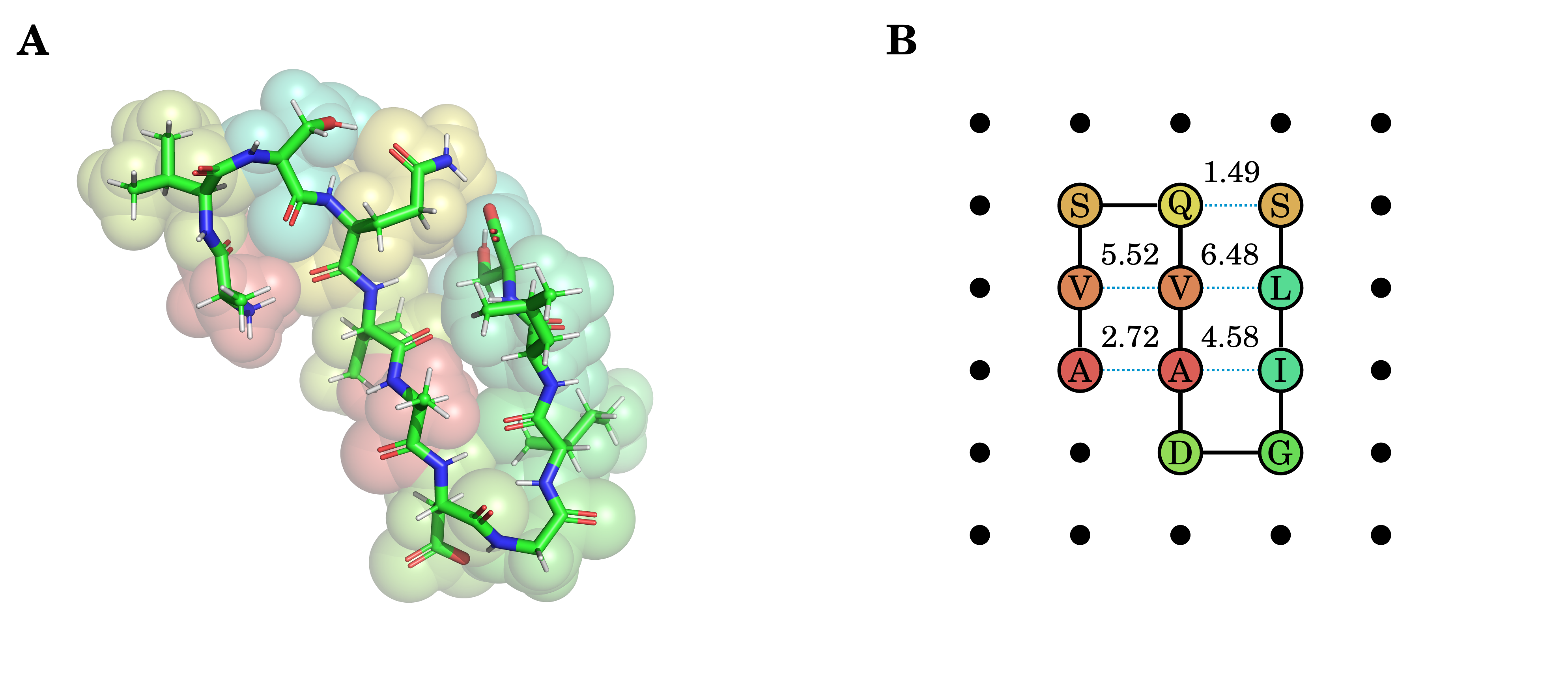}
    \caption{(A) Liquorice representation of a randomly generated short protein (peptide) with sequence AVSQVADGILS. In this depiction, every stick represents a bond between two atoms, and the colour of the corresponding half of the stick identifies the nature of the atom: green is carbon, white is hydrogen, blue is nitrogen and red is oxygen. The spheres that surround the sticks, representing van der Waals volume, have been coloured by residue identity. (B) Lattice model of the peptide in (A). The protein is represented as a self-avoiding walk on a lattice, where every node corresponds to a residue. Amino acids that are distant in the sequence but are spatial neighbours induce complex interactions (represented as dotted blue lines) that stabilise a particular fold. Above the dotted lines, we display the Miyazawa-Jernigan stabilisation energy of the contact.}
    \label{fig:protein-to-lattice}
\end{figure}

Protein chemistry applications of quantum computing have concentrated on a simplified prototype known as the protein lattice model \cite{lau1989hpmodel}, which has been used as a coarse-grained proxy for structure prediction \cite{skolnick2001ab,hoque2009extended} (see Figure \ref{fig:protein-to-lattice}) as well as an invaluable testbed in early theoretical protein folding studies \cite{dill1995principles}. In this model, the protein is described by a self-avoiding walk on a lattice, whose energy is determined by the contacts between adjacent amino acids, and the minimum energy is identified with the biologically active form of the protein. Unfortunately, the problem of finding the protein configuration that minimises the energy is known to be NP-hard \cite{hart1997robust,berger1998protein}. In the context of quantum computing, several encodings (\textit{i.e.} instructions to map the problem to a Hamiltonian operator and the solutions to a binary string) have been proposed \cite{perdomo2008folding,babbush2012encoding,babej2018coarse}, some of which have also been tested experimentally in D-Wave processors \cite{perdomo2012experiment,babej2018coarse}. Recent work has attempted extensions of the protein lattice model \cite{robert2019resource}, and even off-lattice models \cite{mulligan2019designing,casares2021qfold}. Although multiple algorithmic approaches have been suggested, there is not, to our knowledge, any numerical or analytical study establishing the computational scaling of quantum annealing for protein folding applications.

Protein sequences are constrained to a small range of hundreds, or at most thousands of amino acids. This idiosyncrasy renders considerations of asymptotic complexity, a typical focus of the quantum algorithms literature, inadequate for a practical evaluation of usefulness. The median length of a human protein is 375 amino acids long \cite{protlength}, while the median clinical target is about 414 amino acids long \cite{druglength} -- and the effective length may be made smaller by considering independently-folding domains, an ubiquitous strategy in computational structure prediction pipelines. Hence, most interesting problems could be encoded with just a few hundred qubits. If there exists a relative, not necessarily asymptotic, speedup with respect to classical heuristics, this will be of interest to both basic research in molecular biology and the pharmaceutical and biotechnological industries, which dedicate significant resources to structural protein studies. Establishing the scaling of quantum annealers in problems of modest size is a crucial baseline for further exploration as near-term quantum annealers capable of addressing larger problems are made available \cite{dwave2020advantage}.

In this article, we present an extensive numerical study of protein lattice folding in idealised, error-free, closed-system quantum annealing, such as might be achievable in future devices with long coherence times in the presence of error correction \cite{lidar2008towards,pudenz2014error}, or with fault-tolerant universal quantum computers employing Hamiltonian simulation \cite{lloyd1996universal}. In particular, we have computed the minimum spectral gaps and optimised time-to-solution (TTS) metrics for a large dataset of hard problems. The spectral gaps display a strongly vanishing behaviour, which according to the adiabatic theorem leads one to anticipate a quadratically stronger upper bound in runtime. However, our simulations of unitary evolution reveal a scaling that is several orders of magnitude milder, and that can be optimised further via simple heuristics based on average gap positions.. When compared with classical stochastic optimisation heuristics, we find some evidence of a potential limited quantum speedup for protein lattice folding problems of modest size.

\section{Methodology}
\label{section:methods}

Throughout this article, we use a large dataset of peptide (\textit{i.e.} short protein) sequences, each with a unique global minimum (UGEM), which have been mapped to Ising-like problems with the methods described by Babbush et al. \cite{babbush2012encoding}. Peptide sequences with UGEMs have been shown to display the properties of real proteins, even for small sequences (\textit{e.g.} \cite{dill1995principles}). The dataset is examined using numerical simulations to assess the gap and expected time-to-solution, studied in terms of the relationship between conditions and results, and compared against an off-the-shelf classical algorithm.

\subsection{Benchmark dataset generation}

We examined a total of 29,503 peptide sequences, an approximately equal number of cases in both dimensions (15,173 in 2D and 14,330 in 3D) and lengths (approximately 4,500 cases per length at a given dimension, with the exception of 2D length 7, where it is challenging to generate UGEM cases, and we considered only 1,700 cases). To produce  this dataset, we generated protein sequences by random sampling with replacement of the 20 standard amino acids (ARNDCQEGHILKMFPSTWYV). The states of these instances were enumerated by a brute force algorithm, scoring the energies using the Miyazawa-Jernigan 20-amino acid model \cite{miyazawa1985model}, and all the sequences with two or more non-equivalent minimum energy conformations were rejected.

These sequences {were mapped} to an algebraic expression representing the couplings between individual spins in a programmable Ising model. This expression is often known as a Polynomial Unconstrained Binary Optimisation (PUBO) \cite{dwave2019handbook}. We employ  the \textit{turn encoding} approach by Babbush et al. \cite{babbush2012encoding}, which displays the highest reported efficiency in the number of qubits. {To perform the mapping in practice, we developed a} modified version of \textit{SymEngine} \cite{symengine}, a computer algebra system (CAS) {written} in C\texttt{++}{, which} exploits the idempotency of binary variables leading to up to a five orders of magnitude speedup. {We consider protein Hamiltonians with 10 (6 aa peptide in 2D) to 21 qubits (8 aa peptide in 3D), meaning Hilbert spaces of size 1,024 to 2,097,152.}

\subsection{Numerical simulations}

\subsubsection{Representation of the Hamiltonian}

Every quantum annealing process considered in this article may be represented in the following general form:

\begin{equation}
\label{eq:Hamiltonian-evolution}
H(s) = \left(1-s\right)H_\mathrm{start} + sH_\mathrm{protein} + \lambda s(1-s)H_\mathrm{catalyst}
\end{equation}

where $H_\mathrm{protein}$ is the Hamiltonian expressing a given protein lattice problem,

\begin{equation}
\label{eq:start-Hamiltonian}
H_\mathrm{start} = \frac{1}{2}\sum_{i=1}^N \left(I-\sigma^x_i\right)
\end{equation}

and $H_\mathrm{catalyst}$ is one of the following:

\begin{equation}
\label{eq:nonstoquastic-catalyst}
H_\mathrm{nonstq} = \sum_{i=1}^N\sum_{j=1}^N \sigma^x_i\sigma^x_j
\end{equation}
\begin{equation}
\label{eq:stoquastic-catalyst}
H_\mathrm{stq} = \sum_{i=1}^N \sigma^x_i
\end{equation}

In these equations, $N$ is the number of qubits; $\lambda$ is the strength of the catalyst; $s$ is the annealing progress, defined in the interval $[0, 1]$, and generated by some interpolation function $s=f(t/T)$; $I$ is the identity operator in the $N$-spin space \textit{i.e.} $I=I_1\otimes I_2\otimes\mathellipsis I_N$; and $\sigma^x_i$ is the Pauli X matrix applied to the $i$th spin in $N$-spin space, \textit{i.e.} $\sigma^x_i=\left(\bigotimes_{k=1}^{i-1} I \right)\otimes \sigma_x \otimes\left( \bigotimes_{k=i+1}^N I\right)$. 

In the computational basis, $H_\mathrm{protein}$ can be represented as a diagonal matrix, and both $H_\mathrm{start}$ and $H_\mathrm{stq}$ will be matrices with off-diagonal elements in defined positions. It is easy to prove that the number of null elements in a linear combination of these Hamiltonian matrices grows as $\mathcal{O}(2^N)$, hence significant savings in memory and processing time can be made by exploiting the sparsity of the matrix representation. In contrast, however, $H_\mathrm{nonstq}$ has a different sparsity pattern and can only be simulated at a much higher computational cost. Our numerical simulations relied on the {\tt PETSc} library \cite{abhyankar2018petsc,balay2019petsc}, where the Hamiltonians were represented in the compressed sparse row (CSR) or Yale format.

\color{black}
\subsubsection{Gap evaluation}

We estimated the minimum spectral gap between the ground state and the first excited state via numerical diagonalisation of the Hamiltonian matrices, using the Krylov-Schur method \cite{hernandez2007krylov,stewart2002krylov} as implemented in the \texttt{SLEPc} package \cite{hernandez2005slepc,hernandez2007krylov}. We computed the eigenvalues in increments of $\Delta s=0.1$, and interpolated the results using cubic splines. The gap was found as the minimum energy difference between every curve that ran into the ground state at the end of the evolution, and any other curve.

We considered only Hamiltonians without a catalyst term (\textit{i.e.} $\lambda=0$) for the evaluation of the spectral gap.

\subsubsection{Quantum annealing simulation}

We estimated the expected time-to-solution (TTS) by simulating the quantum dynamics of the annealer. We assumed an idealised quantum annealer at zero temperature, in the absence of noise, and with perfect control over couplings, which was simulated by numerical integration of the time-dependent Schr\"odinger equation:

\begin{equation}
\frac{d\ket{\Psi}}{dt} = -iH(t)\ket{\Psi}
\end{equation}

where $H(t)$ is the time-dependent Hamiltonian defined in equation \ref{eq:Hamiltonian-evolution}. We integrated this equation using the Runge-Kutta 5\textsuperscript{th} order method with adaptive timestepping, as implemented in the \texttt{PETSc} package \cite{abhyankar2018petsc,balay2019petsc}. Runge-Kutta methods have been previously validated for studying quantum annealing \cite{farhi2000numerical}. At the end of the evolution, the final state $\ket{\Psi(t=T)}$ is a vector of amplitudes $\Psi_i$ whose square modulus $|\Psi_i^2|$ is the probability of measuring a particular binary string in the device.

Several authors have attested the need to use optimal time-to-solution (TTS) metrics to assess the scaling of quantum annealing algorithms, \textit{e.g.} \cite{ronnow2014defining,albash2018demonstration}. We employed the Bayesian optimisation package \texttt{GPyOpt} \cite{gonzalez2016gpyopt} to optimise the annealing time $T$. We defined an optimisation domain between 0.1 and 1,000 a.u., which was considered acceptable after initial exploration{.}

We set up a maximum of 50 iterations for annealing programs involving a single parameter (the sample time), and a maximum of 500 iterations for trajectories including a catalyst, where the strength of the catalyst ($\lambda$) has to be optimised alongside the sample time; in the case of the non-stoquastic catalyst, given the loss of sparsity of the Hamiltonian matrix, the optimisation was stopped after 30 iterations for some of the length 9 peptides. Default parameters were used otherwise.

\subsection{Optimal trajectory design}

We considered the design of tailored trajectories that increase the efficiency of the algorithm. These trajectories were designed by considering the tendency of gaps to appear more often at certain stages of the annealing trajectory (see Figure \ref{fig:gap-distribution}). We developed a heuristic that accounts for the relative probability of encountering a minimum gap, considering the magnitudes of all the gaps found at that position.

Our heuristic is based in dividing the annealing trajectory in small regions ($\Delta s = 0.05$) and estimating the following magnitude for each bin, which we denote ``$R$-score'':

\begin{equation}
R_k = \sum_j\frac{P_{kj}}{f(\Delta_{kj})}
\end{equation}

where $k$ is the index of the bin; the sum extends to all peptides whose minimal gap falls in that bin, indexed by $j$; $P_{kj}$ is the normalised probability that there is a gap at that position (estimated using kernel density estimation with the Epanechnikov kernel and 0.01 bandwidth); $\Delta_{kj}$ is the magnitude of said gap; and $f$ is a function that weighs the magnitude of the gap into the $R$-score. The values of the $R$-score are normalised and used to define a piecewise linear interpolation function that slows down in regions likely to contain a large gap, and anneals at a faster rate otherwise.

We consider two types of optimisation: one employing peptides of all lengths, and a different one designing an optimal program for each length. We also consider several functional forms $f(\Delta_{kj})$ for the R-score: linear ($x$), square root ($\sqrt{x}$) and cubic root ($\sqrt[3]{x}$). Performance comparisons for these functions are reported in Figures \ref{fig:optimal-program} to \ref{fig:length-program-relative}.

\subsection{Classical simulated annealing}

We compared the performance of the quantum annealing to classical simulated annealing, one of the most common optimisation algorithms in computational biology, which is used in many modern structure prediction packages like Rosetta \cite{rosetta}. We employ an off-the-shelf classical simulated annealing subroutine, the {\tt gsl\textunderscore siman.h} module of the GNU Scientific Library \cite{gsl}. This module requires a definition of the optimisation space, and a function that returns the energy of a given configuration. We implemented a simple subroutine to compute the energy of a sequence of lattice moves and, in order to avoid biases deriving from our implementation, we analyse the results in term of Monte Carlo moves instead of times.

The classical simulated annealing subroutine employs several parameters, including the number of trials per step, the number of iterations per temperature, the maximum step size of the random walk and the parameters of the Boltzmann distribution (initial and final temperature, Boltzmann constant and damping factor). All of these parameters, except for the Boltzmann constant (set to 1) and the step size (set to 1), were optimised using Bayesian optimisation in the same way parameters were tuned in numerical simulations of quantum annealing.

\subsection{Statistical analysis of scaling}

We assessed the significance of these results using a statistical analysis. We performed a non-linear least squares fit of our data using the \textit{lmfit} library \cite{newville2016lmfit}. We considered four functional models: polynomial ($x^\alpha$), exponential ($e^{\alpha x}$) square exponential ($e^{\alpha x^2}$) and cubic exponential ($e^{\alpha x^3}$). The function was augmented by a constant $\beta$, which was set to the average value of the dependent variable for a given length. 

We employed three statistical model selection criteria to decide the function that provided the best explanation of the data: the Akaike information criterion (AIC), the Bayesian information criterion (BIC), and the mean squared error (MSE) of the means. The last method was chosen because of the nature of the dataset: the independent variable is discrete, hence the model only provides the prediction for 4 values (3 in 3D). Thus, understanding how the real means differ from the predicted means provides useful information for model selection.

\section{Spectral gap}

\begin{figure*}
    \centering
    \includegraphics[width=\textwidth]{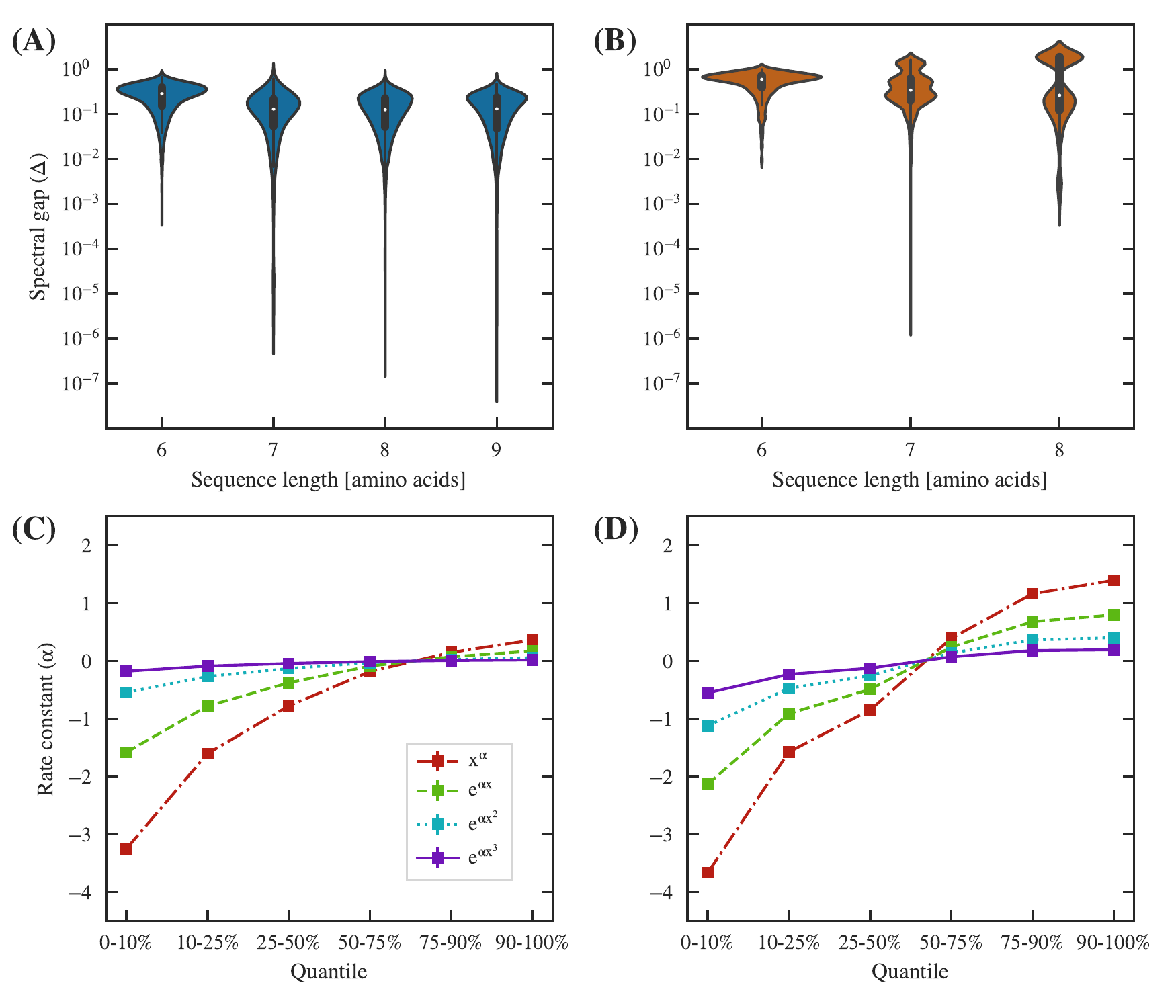}
    \caption{(A, B) Distributions of $\log_{10} \Delta$, the decimal logarithm of the minimum spectral gap between the ground state and first excited state energy levels, for UGEM peptides of different size in 2D (A) and 3D (B). The sequence length is given in number of amino acids (aa). These violin plots employ a Gaussian kernel density estimation method to show a smooth representation of the distribution of data. In the 2D case (A), it is clear that the median gap remains approximately constant, while the worst case gap grows exponentially. In the 3D case (B), a similar effect is observed, although it is made less clear by the particular behaviour of length 8 peptides, which always have their optimal structures arranged in a cube. (C, D) Least-squares fit of subsamples of the data to different functional models. The rate of decay of the minimum spectral gap varies significantly between quantiles.}
    \label{fig:gap_plot}
\end{figure*}

We start our analysis by studying the minimum spectral gap, $\Delta$, between the ground state and the first excited state. It is often stated (based on theoretical arguments) that the runtime of quantum annealing algorithms is proportional to $\mathcal{O}(\Delta^{-2})$ \cite{albash2018adiabatic}. Unfortunately, many problems exhibit an exponentially vanishing gap with increasing problem size \cite{van2001powerful,van2001limits,reichardt2004quantum}, and in particular it is believed that no form of quantum computing is able to efficiently solve NP-complete problems, or at least no such report has withstood scrutiny \cite{farhi2001quantum,albash2018adiabatic}. The distribution of gaps for the protein lattice problem in 2D and 3D, considering a stoquastic process without a catalyst, is shown in Figure \ref{fig:gap_plot}.

The distribution of gaps at a given length resembles a skewed Gaussian distribution: the majority of gaps are concentrated around a narrow center spanning two orders of magnitude, and there are long, thick tails (containing 5-10\% of the data) that spread away several more orders of magnitude to one side. The extent of these tails present{{s}} a severe decrease. In the dataset of 2D peptides, the size of the worst-case gaps decrease{{s}} by five orders of magnitude within a length increase of four amino acids, but only an order of magnitude and a half within the last three lengths considered. Similar results are seen for the 3D peptides. Length 8 three-dimensional peptides show smaller gaps because of their distinct distribution of energetic levels due to cubic symmetry.

This steep decline for the hardest instances is in contrast with the small decrease experienced by an average peptide. The vast majority of the examples populate the area around the median gap, exhibiting a steady but far slower decline. A close examination of the violin plots for the 2D examples also reveals that the position of the peak of the distribution tends to rise as the sequence length increases, and in fact, if we ignore the tails of the distribution, the average gap increases rather than decreases. In interpreting this finding, it is important to keep in mind that our dataset is composed of peptide folding problems that are hard by design, since they have only one ground state solution (plus, in some cases, symmetry-equivalent configurations). These problems are known to be classically very hard \cite{berger1998protein,hart1997robust}. In addition to the lack of structure of NP-hard problems, the proportion of lowest-energy solutions in the solution space is minimal, so randomised algorithms will find it very challenging to find the ground state.

We characterised the scaling of the gap using regression analysis by Maximum Likelihood Estimation (see Section \ref{section:methods} for details). Four functional forms were considered: polynomial, $x^\alpha$, exponential, $e^{\alpha x}$, square exponential, $e^{\alpha x^2}$ and cubic exponential, $e^{\alpha x^3}$. We then employed several standard model selection criteria, detailed in \ref{appendix:statistics}, to select the model that better explains the data. The polynomial model $x^\alpha$, with $\alpha\approx-0.75$ in 2D and $\alpha\approx-0.4$ in 3D, is selected by all criteria, and is significantly better than the second best model.

We also binned the data into different quantiles and repeated the inference, to account for the inhomogeneity of the results (see Tables \ref{table:gap_all_2d} and \ref{table:gap_all_3d} in \ref{appendix:statistics}). In 2D, the polynomial model is consistently selected across all quantiles, and in 3D, there are some quantiles where the exponential and cubic exponential are selected, which is probably due to the limited range of the dataset and the symmetry effect discussed above. We observed a notable variation of the coefficients across quantiles, as depicted in Figures \ref{fig:gap_plot}C and \ref{fig:gap_plot}D. For example, the first quantile, 0-10\%, with $\alpha=-3.23$ in 2D and $\alpha=-3.65$ in 3D for the polynomial model, contains examples whose gaps vanish at a notably larger rate. On the other hand, the two upper quartiles (75-90\% and 90-100\%) display positive $\alpha$ values, showing that the gap actually widens with increasing size. Only a portion of the problem instances exhibits fast gap vanishing.

This data does not allow us to conclude that the gap vanishes polynomially. Our results are reminiscent of a previous study on the Exact Cover problem by Young et al. \cite{young2008size,young2010first}. In that study, it was described that, while the scaling of the spectral gap at small problem sizes is consistent with a polynomial \cite{farhi2001quantum}, at large problem sizes the scaling turns exponential. Similarly, the fraction of problem instances exhibiting small gaps increases at large problem size \cite{young2010first}. We have been unable to study the behaviour of the protein lattice problem at greater sizes, given the large number of qubits and the difficulty of obtaining Hamiltonian expressions beyond 9 amino acids. However, we hypothesise that the protein lattice problem presents exponentially vanishing spectral gaps that will hinder a general polynomial-time solution by quantum annealing for large sizes, while still exhibiting reasonable advantage for problems of modest size, such as may arise in high-throughput peptide discovery experiments.

\section{Simulations of stoquastic dynamics}

The minimum spectral gap imposes an upper bound on the running time of quantum annealing, but in order to understand the behaviour of the process we need to access the evolution of the quantum state during the computational process. We investigate this regime by numerical integration of the Schr\"odinger equation. Since this procedure is costly,  the assessment of our entire dataset of \textit{circa} 30,000 peptides is beyond our resources and, instead, we selected two samples based on spectral gap values. The first sample contains the set of peptides with the smallest gaps (worst-case set), while the second sample is a random selection of peptides (random set). In both cases, each sample contains 100 peptides per chain length of a given dimension, giving a total of 1,400 instances. 

\begin{figure*}[!htbp]
    \centering
    \includegraphics[width=\linewidth]{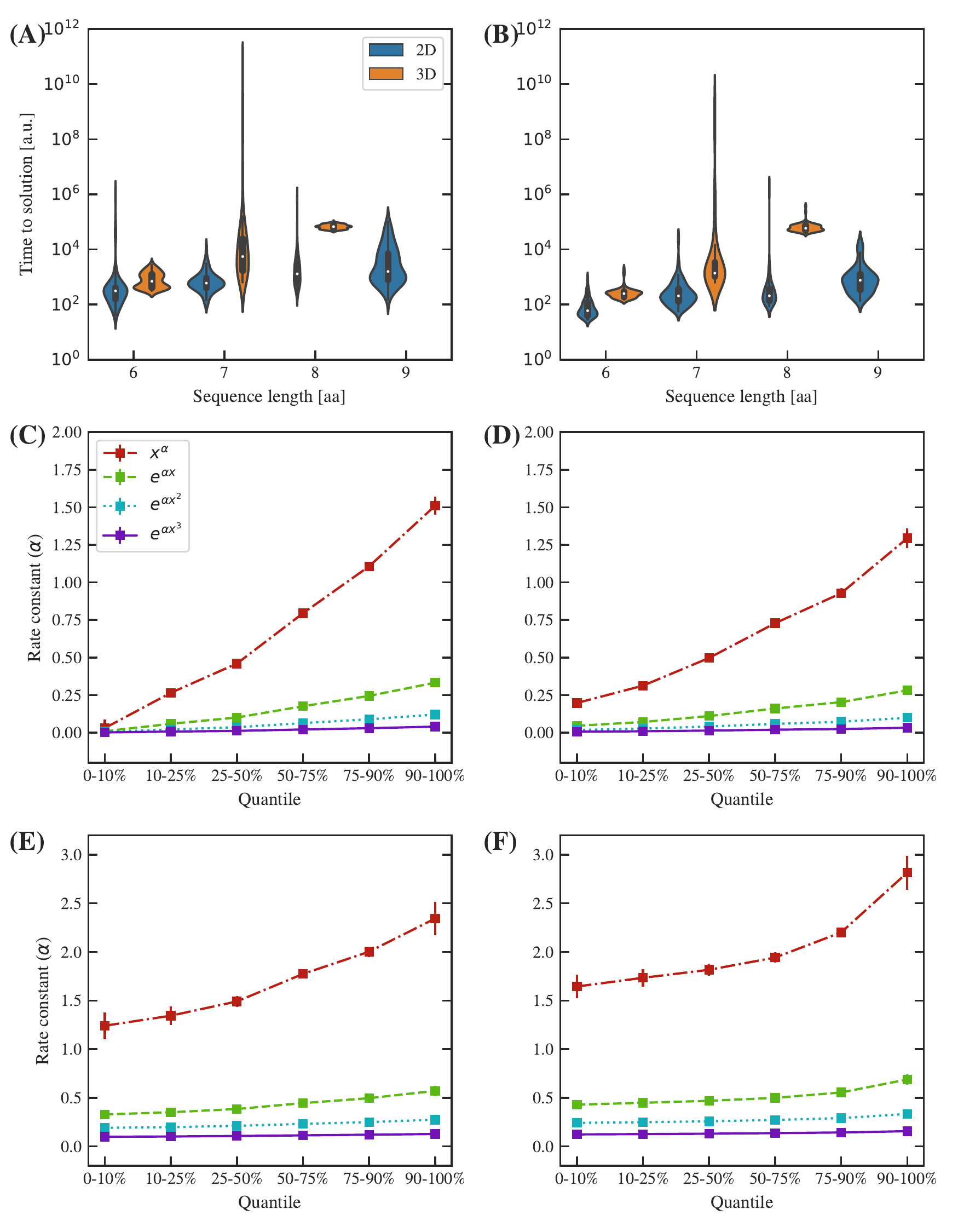}
    \caption{(A, B). Distributions of the expected quantum annealing runtime $T$ of the worst-case (A) and random (B) sets. (C, D). Least-squares fit of subsamples of the data to different functional models for the worst-case (C) and random (D) sets.}
    \label{fig:time_plot}
\end{figure*}

A comparison of this sort requires optimising the annealing time to maximise the probability of success. As described by R{\o}nnow et al. \cite{ronnow2014defining}, a short run can provide a small, but sizeable probability that can be amplified by repetition. In many cases, the repetitions amount to a much shorter runtime {than} a longer, quasi-adiabatic runtime. We employed Bayesian optimisation \cite{shahriari2015taking,gonzalez2016gpyopt} to find the optimal runtime, as detailed in section \ref{section:methods}. The optimised time-to-solution metric, corresponding to the expected runtime to find the correct solution with probability 50\%, is shown in  {Figure \ref{fig:time_plot}A-B}.

The optimisation of the annealing time of an individual run has a significant impact. We simulated a baseline experiment in which the quantum annealer was run for 1000 a. u., and found an average improvement in the expected total runtime of 15 orders of magnitude in 2D and 10 orders of magnitude in 3D {(see Figure \ref{fig:relative-improvement})}. We also find a small, but appreciable difference on the dependence on the gap, as depicted in Table \ref{table:corr_coefs}.

\begin{table}
    \centering
    \begin{tabular}{ccccc}
    \toprule
     & $\rho$ 2D & $\rho$ 3D & $R$ 2D & $R$ 3D \\
    \midrule
    Optimised time & -0.66 & -0.44 & 0.52 & 0.38 \\
    Baseline & -0.73 & -0.34 & 0.62 & 0.30 \\
    \bottomrule
    \end{tabular}
    \caption{Correlation coefficients between expected runtime and gap, for the results obtaining optimising the individual runtime and the results obtained with a fixed runtime of 1,000 a.u. $\rho$ is Spearman's rank correlation coefficient between $\Delta$ and $T$, which describes the monotonic relationship between the two variables ($\rho=+1$ is perfectly monotonic, $\rho=-1$ is perfectly inverse monotonic). $R$ is Pearson's correlation coefficient between $\log_{10}\left(\frac{1}{\Delta^2}\right)$ and $\log_{10}T$}
    \label{table:corr_coefs}
\end{table}

For both the 2D and 3D peptides, the worst-case set containing the smallest gaps does not require significantly longer expected runtimes than the random set. We performed a two-tailed Welch's $t$ test, and found that the random and worst-case sets could not be distinguished ($p$-value 0.34, average $p$-value of subgroup analysis 0.31). In other words, the fact that a problem presents a very small minimum spectral gap does not necessarily indicate it will require a long runtime.  This apparently contradictory statement might be explained by the fact that the $\Delta^{-2}$ scaling in relation to the spectral gap is only valid in the asymptotic limit, and that other terms may govern the dynamics at smaller sizes; and that this rule is above and after all an upper bound, and it is very possible that an instance succeeds even if the annealing is conducted diabatically. In practical terms it is not necessary to guarantee that the final state has a large fidelity with the ground state, but rather to ensure that it has a sizeable amplitude in order to obtain the expected result after enough trials. Another possibility to note is that within the range of problem sizes we inspect there may be only one (or a few) instances of the minimum (or near-minimum) gap occurring during the adiabatic sweep, whereas for large problems a near-minimum gap may occur at multiple points. In addition, the decrease by five orders of magnitude in 2D gaps discussed earlier is also markedly absent from Figures \ref{fig:time_plot}A-B

We performed a scaling analysis identical to the previous section, finding that, for all cases, either the exponential model is selected over the polynomial, or there is not a significant difference between both of them (see Tables \ref{table:time_bad_2d} to \ref{table:time_random_3d} in \ref{appendix:statistics}). In particular, in 2D the polynomial model $x^\alpha$ (with $\alpha\approx0.65$) and the exponential model $e^{\alpha x}$ (with $\alpha\approx0.15$ cannot be separated. In 3D, an exponential model $e^{\alpha x}$ with $\alpha\approx0.45$ is selected with high significance.

Our findings suggest that the protein lattice problem is not as severely affected by the vanishing of the spectral gap, as might have been expected. Problems with gaps smaller than $10^{-2}$ a.u. (and down to $10^{-8}$ a.u.) do not take significantly longer than problems with a median gap of 0.22 a.u. Moreover, {our results hint at} an exponential scaling, albeit with a {potentially} small rate constant. This analysis suggests that this quantum annealing application has a milder scaling than previously expected.

\section{Improvement of the annealing pathway}

In the previous section, we found that optimising the sample size can dramatically enhance the performance of quantum annealing, often by several orders of magnitude. This raises the question of whether other simple changes may be able to similarly increase efficiency. In this section we explore several approaches that deliver increasing improvements to the performance of the algorithm. 

We first considered whether the linear interpolation function is the most appropriate choice, or if conversely a non-linear interpolation function increases the efficiency of the algorithm. As a first approach, we experimented with three alternative functions: quadratic ($x^2$), cubic ($x^3$) and, in order to include a rapidly changing schedule, sigmoid ($1/(1+e^z)$). We optimised these functions following the same procedure as in the previous section. The results are summarised in Figures \ref{fig:interpolation-functions} and \ref{fig:interpolation-functions-relative}.

Our results suggest that none of the functions considered is able to deliver a significant improvement. The quadratic function displays only minor deviations, and the cubic and sigmoid functions lead to markedly worse performance 99\% of the cases. Moreover, when the runtime was reduced, the magnitude of the change was much smaller (roughly 80\% of the original time to solution) than when the runtime was slowed down (140-150\% in the cubic case and 230-300\% in the sigmoid). This supports our initial choice of the linear annealing schedule to establish general trends.

We then considered the introduction of a non-stoquastic catalyst, which has been considered as a potential technique to improve quantum annealing \cite{hormozi2017nonstoquastic,albash2019role}. As detailed in the Methods section, this catalyst incorporates terms $\sigma^x_i\sigma^x_j$, and is multiplied by a tunable parameter $\lambda$ that was optimised alongside the sample time. Introducing a catalyst of this form alters the sparsity of the Hamiltonian matrix that our code was optimised for, and therefore we could only consider one of the interpolation functions. This is adequate since our previous examination reveals that the choice of interpolation function has a small impact on efficiency. The results of these simulations are reported in Figures \ref{fig:non-stoquasticityXX} and \ref{fig:non-stoquasticity-relativeXX}.

The introduction of a non-stoquastic catalyst improves the runtime in about 5\% of the worst-case examples, showing a notable average slowdown in the random sample. The magnitude of the deterioration seems to increase with the size of the sequence, while the proportion of sequences that are improved, as well as the magnitude of said improvement, decrease with the size of the sequence. These results suggest that while the non-stoquastic catalyst may well be useful in a fraction of the hardest problems, it is not a general solution. Incidentally, our results agree with those in a recent article by Crosson et al. \cite{crosson2020designing}, providing significant analytical and numerical evidence that, in general, stoquastic Hamiltonians are more adept at optimisation than their non-stoquastic counterparts, while allowing however for specific, cleverly engineered non-stoquastic catalysts to provide excellent (sometimes, even exponential) speedups.

The theoretical arguments in \cite{crosson2020designing} led us to consider a \textit{stoquastic} catalyst that perturbs the annealing trajectory while maintaining stoquasticity. We conceived a catalyst that rotates the transverse field throughout annealing with a sum of terms $\sigma^x_i$. Incidentally, this choice of rotation axis preserves the sparse structure of the matrix and enables rapid simulation. As in the previous case, we introduced a tunable parameter $\lambda$ that was optimised alongside the sample time. 

The stoquastic catalyst also improves the runtime in about 5\% of the examples. In this case, however, both the proportion of sequences and the magnitude of the improvement seem to increase with the size of the sequence, and while the improvement is generally more notable in the worst-case dataset than in the random dataset, both present some sort of improvement. The magnitude of the deterioration is significantly smaller than in the non-stoquastic case, and is smaller than an order of magnitude, compared to the 5-6 orders of magnitude introduced before. These results may indicate that a stoquastic catalyst is generally preferable to a non-stoquastic one.

After exploring these alternatives, we considered the design of \textit{tailored} annealing trajectories: schedules designed to slow down near the spectral gap, and therefore reduce the ratio of excitation to local minima. This is motivated by the tendency of spectral gaps to concentrate around certain particular positions (see Figure \ref{fig:gap-distribution}). In practice, a tailored trajectory is simply a piecewise linear function, defined on equally spaced bins, where the slope of the function at a given bin is inversely proportional to the probability of finding a gap of certain magnitude. We report the heuristic employed to design these schedules in the Methods section. The results of these experiments are summarised in Figures \ref{fig:optimal-program} and \ref{fig:optimal-program-relative}.

Tailored trajectories\footnote{Here we refer to tailored trajectories employing the cubic root function in the $R$-score (see Section \ref{section:methods}), which are the best performing in our benchmark.} display better results than any of the previous approaches. Nearly half of the sequences in the random dataset, and about 80\% of the worst-case dataset, experience some improvement. The magnitude of the speedup is also significantly larger, with many cases achieving accelerations between 10x and 100x, and we observe an increasing trend where longer peptides improve more than shorter ones. Unfortunately, this technique has a significant caveat: that, when the strategy fails, the deterioration can be far worse ($10^4$ to $10^6$ slow down in some cases).

A primary reason why the tailored approach can result in such slow annealing processes is because our heuristic has to balance the positions where there is a high probability of encountering the gap, and the magnitude of said energy difference. In some cases, particularly in the longer sequences, this balance is failing. A potential solution is to employ the distribution of gaps for every peptide length, instead of considering all the data at once. We therefore designed one annealing schedule per peptide length, and simulated the results as in the previous sections. The results of this experiment are reported in Figures \ref{fig:length-program} and \ref{fig:length-program-relative}.

This approach is unsurprisingly the most successful of all, improving 60\% of the sequences in the random dataset, and an impressive 95\% of the sequences in the bad dataset. The magnitude of the improvement is similar, although the average improvement is higher; and we do not see the significant deterioration observed in the previous approach. This results suggest a potential strategy to optimise quantum annealing further.

In practical applications, the position of the gap will not be readily available. However, we have observed that the position of the gaps can be estimated by running short annealing runs with a sudden change at different parts of the schedule. We considered a piecewise linear function, similar to the one used in the tailored trajectories, but changing the slope only at the point where we probe for the gap. When applied to our dataset, this method shows some correlation (full dataset: $R=0.17$, $p=10^{-6}$; worst-case dataset: $R=0.27$, $p=5\cdot 10^{-8}$), suggesting that more involved programs may be able to produce reasonable estimations of the gap position.

The results in this section suggest that simple engineering of the Hamiltonians and trajectories involved in the annealing process can deliver a substantial increase in performance. In this sense, the results presented earlier represent a ``sensible baseline'' of what is achievable with quantum annealing; and the experiments described in this section propose a pathway to improve the performance further. This also allows us to postulate that further experimentation in this domain may deliver better scalings (in relative terms, but potentially also in complexity) than previously reported.

\section{Comparison with simulated annealing}

We have observed that quantum annealing requires exponentially growing runtimes to find the ground state of a protein lattice model, even in the small range of lengths explored in our dataset (see Figure \ref{fig:time_plot}). Assuming this remains true of larger systems, as one would expect, it precludes an exponential speedup, since enumerating all possible conformations of a lattice model has the same asymptotic complexity. However, an algorithm which scales significantly better, that is, with a far smaller exponential rate $\alpha$ than the classical case could still be useful for practical applications. 

\begin{figure*}
    \centering
    \includegraphics[width=\linewidth]{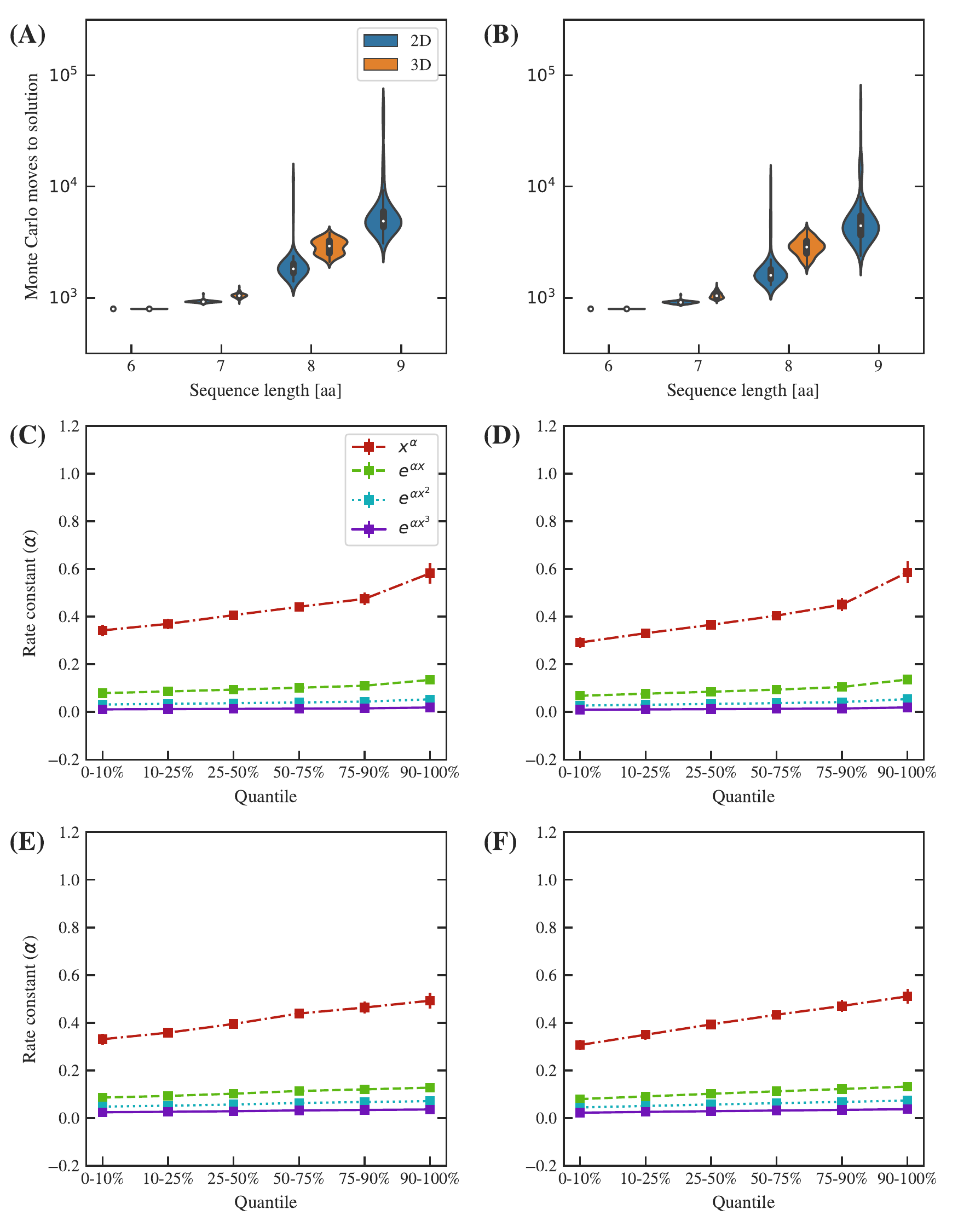}
    \caption{(A, B). Distributions of the expected classical simulated annealing number of Monte Carlo moves $N$ of the worst-case (A) and random (B) sets. (C, D). Least-squares fit of subsamples of the data to different functional models for the worst-case (C) and random (D) sets.}
    \label{fig:sa_plot}
\end{figure*}

In this section, we compare quantum annealing with classical simulated annealing using the data displayed in Figure \ref{fig:sa_plot}. Unlike other comparisons of quantum annealing and classical simulated annealing (\textit{e.g.} \cite{albash2018demonstration}), we have not constrained the classical approach to solve a problem in the Ising form. More importantly, we consider a NP-hard problem, as opposed to previous work by Albash et al. \cite{albash2018demonstration} that considered simpler problems.

The distributions presented in Figure \ref{fig:sa_plot}A and B show a rapidly growing number of {Monte Carlo moves}. Visually, the runtime appears to display worse-than-exponential growth. Our model comparison analysis (see Tables \ref{table:sa_bad_2d} to \ref{table:sa_random_3d} in \ref{appendix:statistics}) finds that, in all cases, the model fits to a square exponential $e^{\alpha x^2}$ with a high level of significance (and this behaviour is reproduced at every quantile). 

There are some theoretical arguments (e.g. \cite{rajasekaran1990convergence}) which conjecture that simulated annealing converges in exponential time. The square exponential fit found by our statistical analysis could be an artifact of parameter optimisation (note that we optimise four parameters for a single simulated annealing run, as opposed to only one in quantum annealing). This anticipates that quantum annealing provides a better scaling. Our results are made stronger by the fact that in this analysis we have considered only the { number of Monte Carlo moves, i.e. the} number of evaluations of the energy function{.}{ The cost of evaluating this energy is approximately quadratic on the size of the peptide, and the actual performance will be slightly worse than as depicted in Figure \ref{fig:sa_plot}. Of course, since this cost is polynomial, there will be no changes to the complexity of the algorithm, albeit the practical scaling will be worse.}

These findings suggest that quantum annealing has an improved performance over simulated annealing. Over the system sizes we examined, the runtime of quantum annealing scales approximately exponentially, while simulated annealing shows a rapidly growing function that fits better to a {square} exponential.

\color{black}
\section{Discussion}
\label{section:conclusions}

In this article, we have presented a numerical investigation of quantum annealing applied to protein lattice models. We have considered nearly 30,000 protein sequences, each with a unique global energy minimum, which represent realistic protein problems displaying a folded state, but are also high difficulty instances of a NP-hard problem.

We first turned our attention to the minimum spectral gap, a quantity connected theoretically to the runtime of a perfect annealer. We have observed that the gap for these protein sequences can decrease quickly in magnitude, although the scaling appears to be polynomial in the range of sizes considered. The polynomial scaling was confirmed by several statistical selection criteria (detailed in the Methods section and \ref{appendix:statistics}), although comparison with prior results reported in the literature leads us to hypothesise that the gap vanishes exponentially. We have also observed that the worst cases decrease by five orders of magnitude between 6 and 9 amino acids. This numerical evidence shows that adiabatic evolution of the computer, where the probability of success nears 100\%, will require rapidly growing runtimes and likely be infeasible for worst-case problems.

We then considered optimal annealing runs, where the computer is run for a shorter time to produce a small, but sizeable probability of success that is amplified by repetition. We have established that this runtime grows exponentially with peptide length, although the rate of growth was far smaller than the gap analysis suggested. The scaling was found to be approximately equal to $e^{0.15L}$ for 2D examples and approximately  $e^{0.75L}$ for 3D examples within the range of problem sizes studied. We also found statistical evidence that peptides with very small gaps are not significantly harder than average cases, and that the exponential rates are almost identical for these two datasets.

{ We experimented with the parameters of the annealing process in pursuit of strategies to increase efficiency. We considered non-linear interpolation functions, non-stoquastic catalysts, and two adaptive programs to slow down the annealing schedule near regions where a potential gap is expected. Our results demonstrate that the last approach is able to deliver 10-100x speedups. Furthermore, we observe a trend of increasing relative improvements as the size increases: the vast majority of the peptides in the worst-case dataset were improved by this strategy, suggesting that potential decreases in performance can be addressed via careful engineering. We also suggest a method to estimate the gap position when this is not be readily available.}

A comparison with classical simulated annealing on our dataset shows that the quantum annealing approach is preferable. Statistical modelling seems to suggest that the scaling of simulated annealing fits best to a square exponential, $e^{\alpha x^2}$ although theoretical arguments lead us to expect this behaviour to become exponential with a large rate as problem size increases. This implies that for large peptide sizes, a quantum annealer may take significantly less time than a classical machine running a stochastic algorithm. 

One of the reasons why quantum annealing may prove useful for protein folding and structure prediction is the limited size of interesting problems. More than half the structures deposited in the Protein DataBank contain fewer than 500 residues, and 80\% of the domains in the CATH database \cite{pearl2003cath} are smaller than 200 residues. Similarly, { the median length of a human protein is 375 amino acids long \cite{protlength}}. Even if the scaling of quantum annealing is exponential, as long as the exponential rate is low enough to fold small proteins or domains in a timely fashion, this approach will be useful for a multitude of practical problems.

There are other advantages to quantum annealing that could be explored further. When the algorithm fails, the system has been excited to a higher energy state, but although this will only be a local minimum, it may still be useful. For example, if this result is used in a bottom-up approach to explore the conformational space of a protein, it may still be a good starting point for more complex simulations. In contrast, classical simulated annealing is not guaranteed to provide a solution that is close to the global minimum. Future work will investigate this fact, and aim to establish ``crossover'' points where the scaling worsens.

We believe these results suggest that annealing approaches to quantum optimisation may be a powerful heuristic approach to solve the protein lattice problem, whether in specialised processors \cite{dwave2020advantage}, near-term digital quantum computers \cite{chen2020demonstration} or otherwise. The difficulties of the quantum approach are shared by the classical simulated annealing approach, while the scaling is better. Moreover, even in cases where the annealing approach fails, it can provide solutions that despite not being equivalent to the global minimum, are very close to it, and may be useful in a subsequent refinement procedure. These findings offer encouragement for further research in quantum protein lattice folding and other hybrid quantum-classical algorithms for protein structure prediction.

\section*{Data availability}
The data that support the findings of this study are available from the corresponding author upon reasonable request.

\section*{Code availability}
The code that supports the findings of this study will be made available after acceptance at \url{https://github.com/couteiral/proteinham}.

\section*{Acknowledgements}
C. O. would like to thank F. Hoffmann-La Roche, UCB and the UK National Quantum Technologies Programme for financial support through an EPSRC studentship (EP/M013243/1). G. M. M. thanks the EPSRC and MRC for support via EP/L016044/1 and EP/S024093/1. S. C. B. acknowledges support from the EU Flagship project AQTION, the NQIT Hub (EP/M013243/1) and the QCS Hub. The authors would like to thank Daniel Nissley and Niel de Beaudrap for useful discussions.

\appendix
\beginsupplement

\FloatBarrier
\section{Additional figures}
\pagebreak
\FloatBarrier

\begin{figure}
    \centering
    \includegraphics[width=\linewidth]{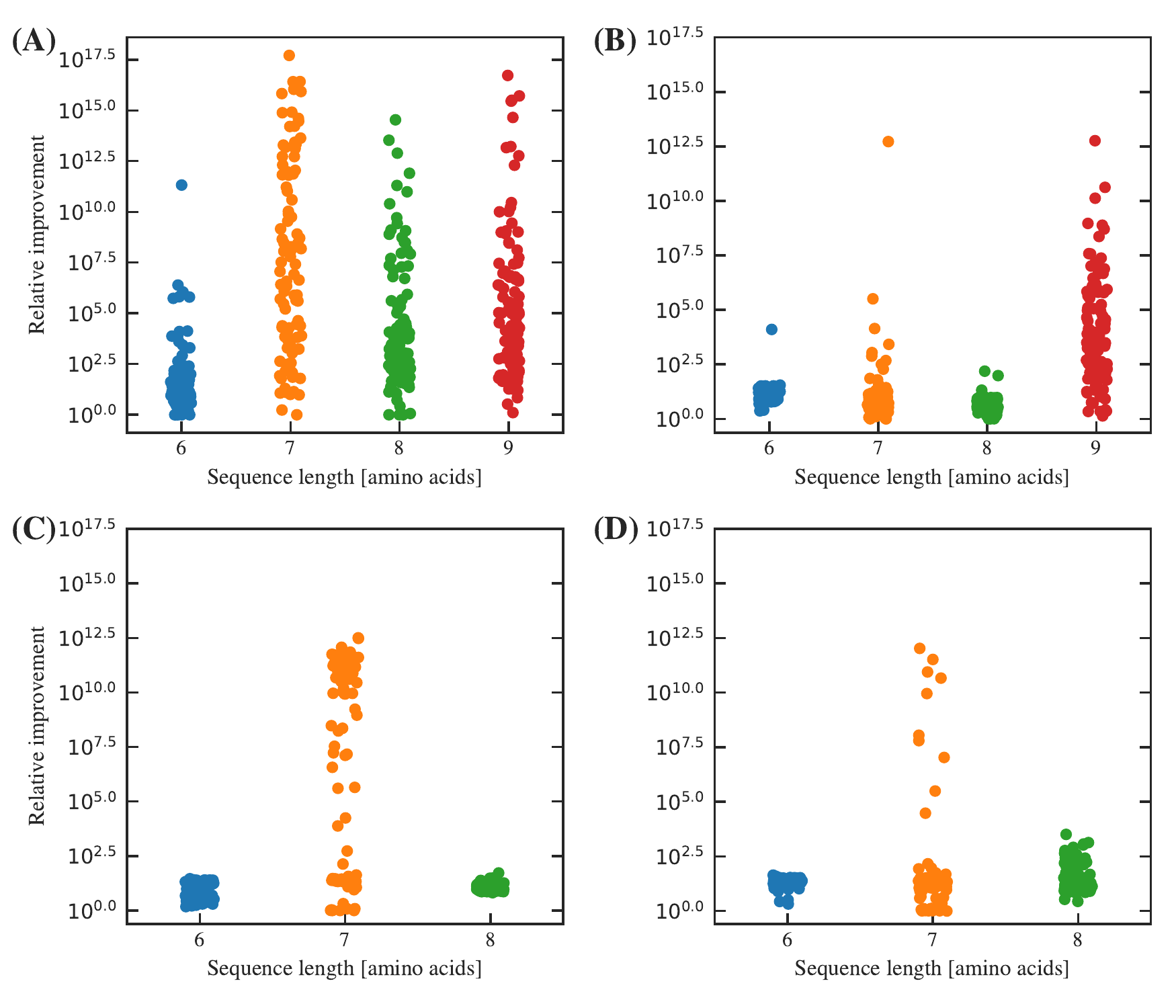}
    \caption{{ Distribution of the relative improvement in the time to solution metric, comparing a baseline 1000 a.u. quantum annealing procedure with an optimised sample time, for (A) the worst-case dataset in 2D, (B) the random dataset in 2D, (C) the worst-case dataset in 3D and (D) the random dataset in 3D.}}
    \label{fig:relative-improvement}
\end{figure}

\begin{figure}
    \centering
    \includegraphics[width=\linewidth]{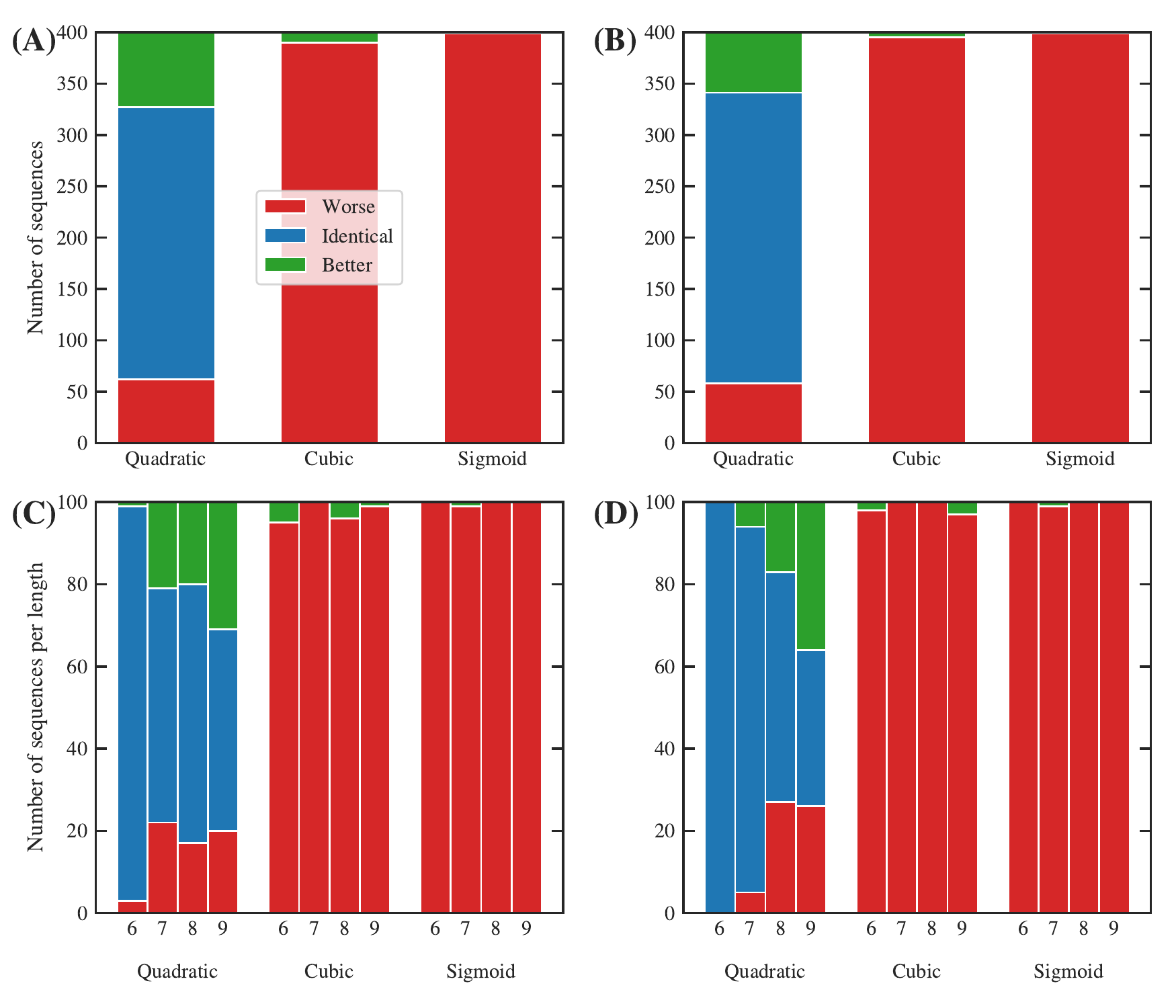}
    \caption{{ Variation of the annealing process using alternative interpolation functions with respect to the linear baseline. We consider quadratic ($x^2$), cubic ($x^3$) and sigmoid ($1/(1+e^z)$) interpolations. (A, B) Proportion of solutions with a worse (red), identical (blue) or better (green) expected runtime than the baseline, for the worst-case (A) and random (B) datasets. (C, D) Proportions classified by length for the worst-case (A) and random (B) datasets. The quadratic function shows scarce deviation from the baseline, while both the cubic and sigmoid functions result in worse performances for all but a few cases.}}
    \label{fig:interpolation-functions}
\end{figure}

\begin{figure}
    \centering
    \includegraphics[width=\linewidth]{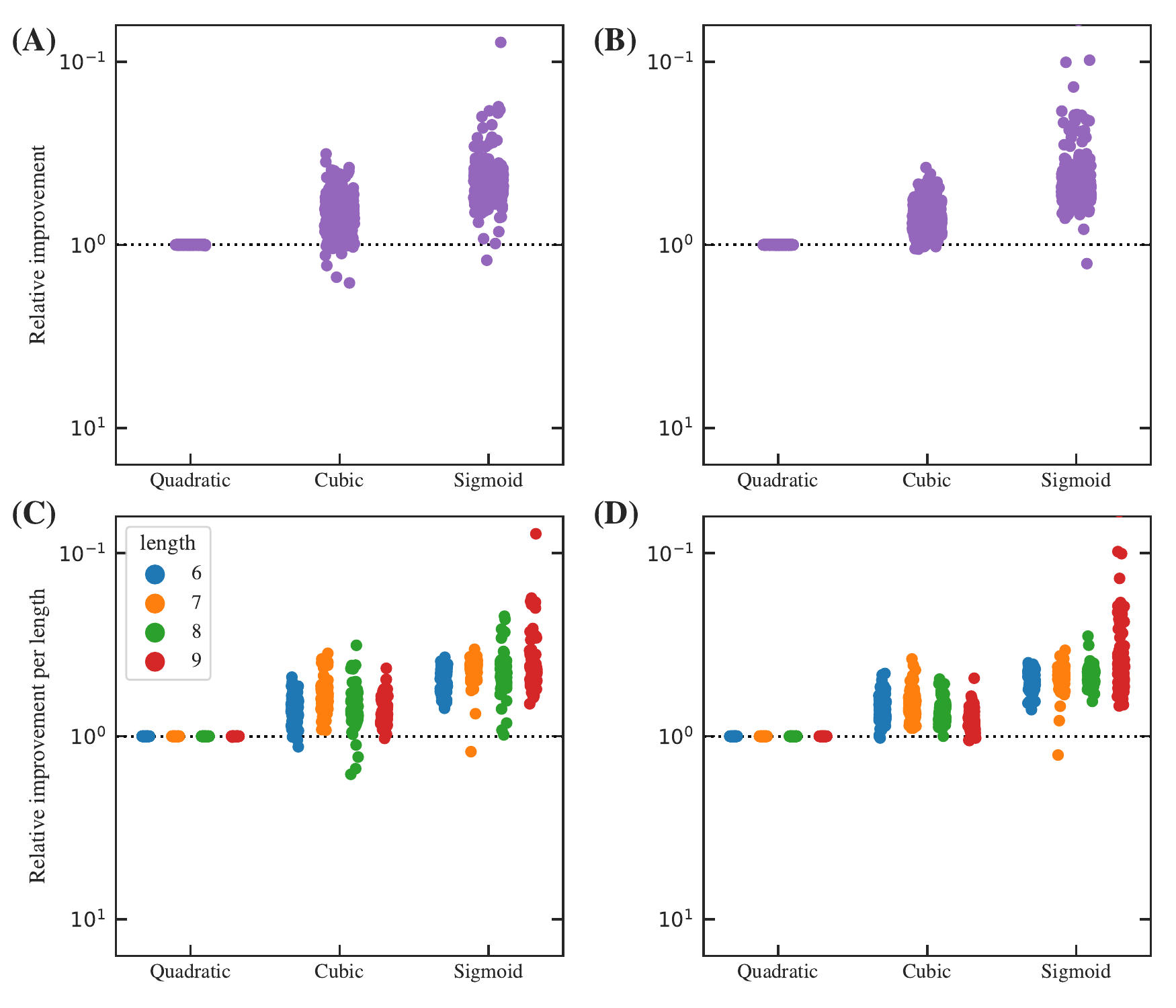}
    \caption{{ Relative improvement of the annealing process using alternative interpolation functions with respect to the linear baseline. We consider quadratic ($x^2$), cubic ($x^3$) and sigmoid ($1/(1+e^z)$) interpolations. Relative improvement of the expected time to solution time for the worst-case (A) and random (B) datasets. (C, D) Relative improvement classified by length for the worst-case (C) and random (D) datasets. Relative improvement here is defined as the ratio between the expected time to solution of the optimised non-stoquastic run and the stoquastic baseline: values under 1.0 (dotted line) indicate worse performance, while values over 1.0 are improved by the choice of interpolation function. We observe that most choices of interpolation functions show worse performance than linear interpolation.}}
    \label{fig:interpolation-functions-relative}
\end{figure}

\begin{figure}
    \centering
    \includegraphics[width=\linewidth]{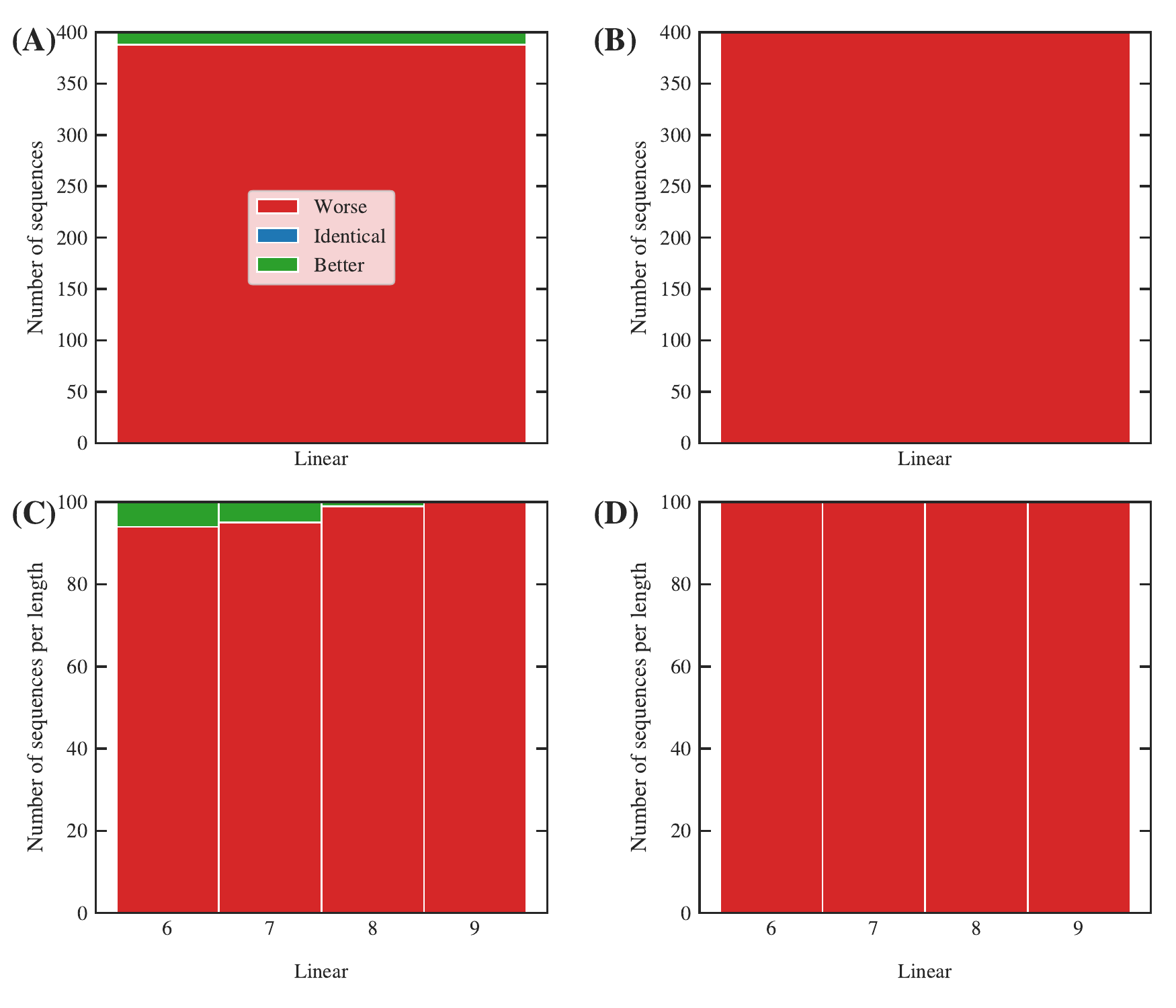}
    \caption{{ Variation of the annealing process with a non-stoquastic catalyst.  (A, B) Proportion of solutions with a worse (red), identical (blue) or better (green) expected runtime than the baseline, for the worst-case (A) and random (B) datasets. (C, D) Proportions classified by length for the worst-case (C) and random (D) datasets. The non-stoquastic catalyst improves the annealing process in a reduced proportion of the worst-case examples, negatively affecting the performance of the rest.}}
    \label{fig:non-stoquasticityXX}
\end{figure}

\begin{figure}
    \centering
    \includegraphics[width=\linewidth]{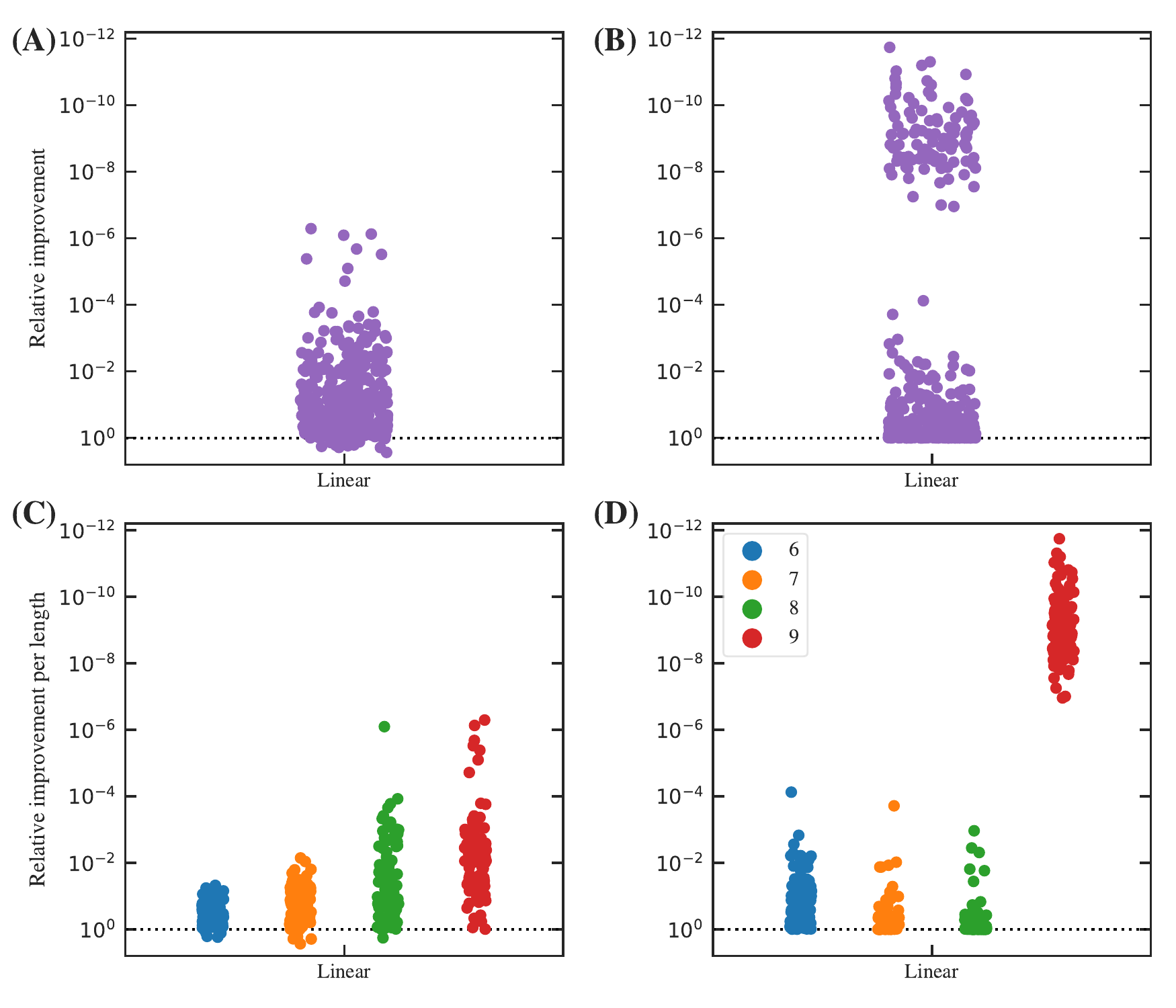}
    \caption{{Relative improvement of the annealing process using a non-stoquastic catalyst, with respect to the stoquastic baseline. (A, B) Relative improvement of the expected time to solution time for the worst-case (A) and random (B) datasets. (C, D) Relative improvement classified by length for the worst-case (C) and random (D) datasets. Relative improvement here is defined as the ratio between the expected time to solution of the optimised non-stoquastic run and the stoquastic baseline: values under 1.0 (dotted line) indicate worse performance, while values over 1.0 are improved by the non-stoquastic catalyst. Our results suggest that the non-stoquastic catalyst notably worsens the performance of quantum annealing in virtually all cases, with a modest improvement in a fraction of the worst-case examples. Note that some of the results for size 9 peptides had to be restricted to a smaller set of iterations due to increased computational burden, which may explain why size 9 random peptides seem to perform significantly worse than the trend would suggest.}}
    \label{fig:non-stoquasticity-relativeXX}
\end{figure}

\begin{figure}
    \centering
    \includegraphics[width=\linewidth]{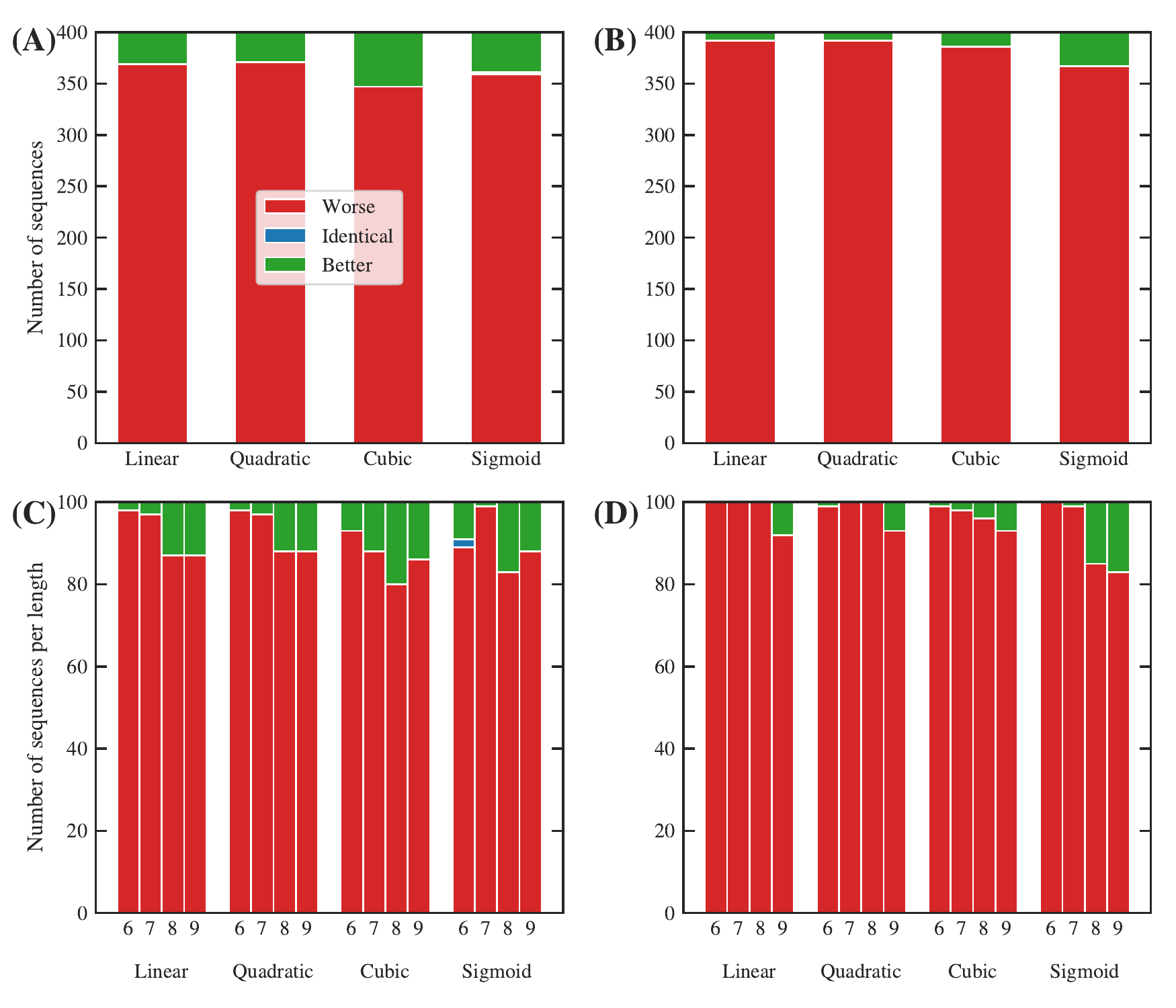}
    \caption{{ Variation of the annealing process with a stoquastic catalyst. We consider linear ($x$), quadratic ($x^2$), cubic ($x^3$) and sigmoid ($1/(1+e^z)$) interpolation functions for the annealing schedule. (A, B) Proportion of solutions with a worse (red), identical (blue) or better (green) expected runtime than the baseline, for the worst-case (A) and random (B) datasets. (C, D) Proportions classified by length for the worst-case (C) and random (D) datasets. The non-stoquastic catalyst improves the annealing process in a reduced proportion of the cases, negatively affecting the performance of the rest.}}
    \label{fig:non-stoquasticity}
\end{figure}

\begin{figure}
    \centering
    \includegraphics[width=\linewidth]{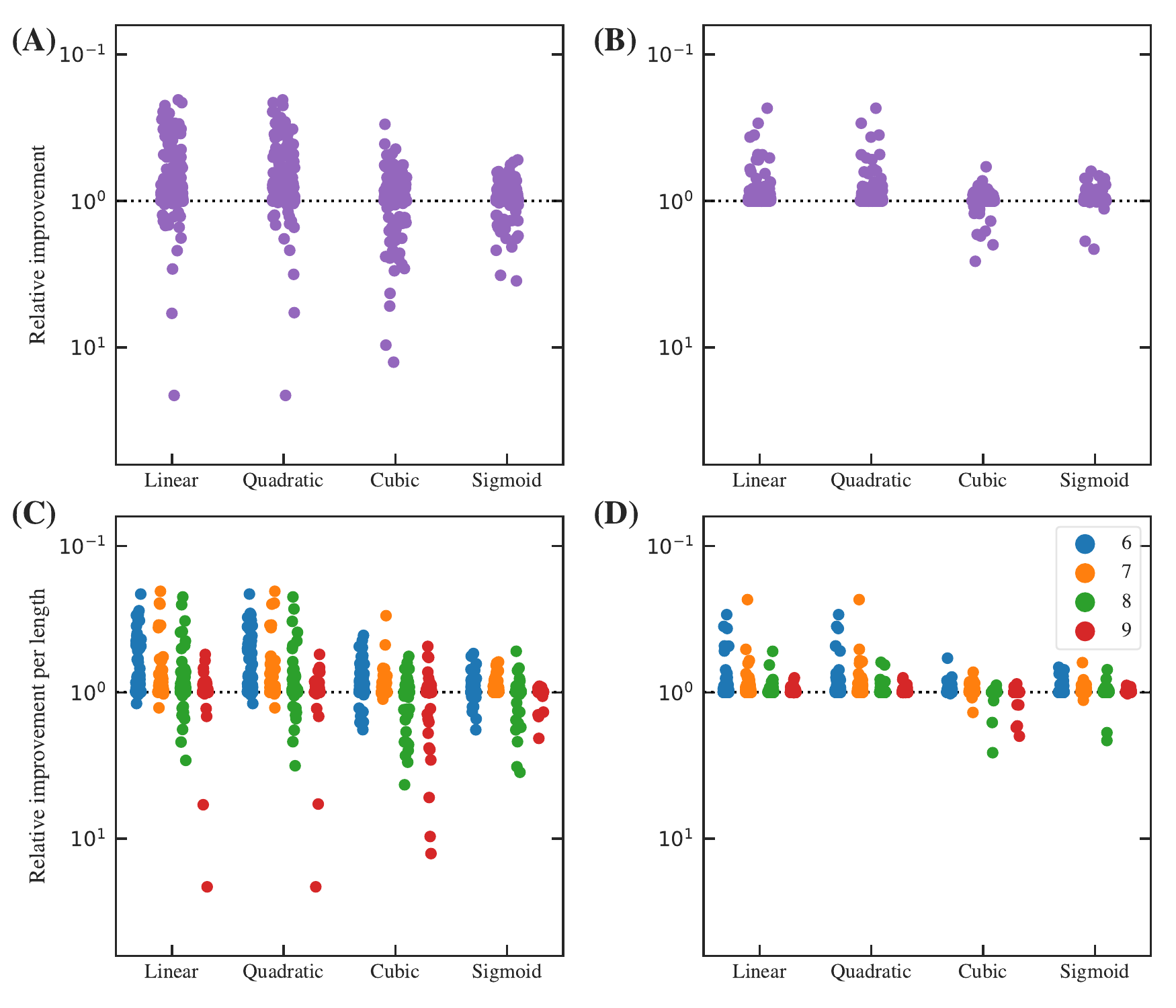}
    \caption{{ Relative improvement of the annealing process using a stoquastic catalyst, with respect to the non-catalysed baseline. We consider linear ($x$), quadratic ($x^2$), cubic ($x^3$) and sigmoid ($1/(1+e^z)$) interpolation functions for the annealing schedule. (A, B) Relative improvement of the expected time to solution time for the worst-case (A) and random (B) datasets. (C, D) Relative improvement classified by length for the worst-case (C) and random (D) datasets. Relative improvement here is defined as the ratio between the expected time to solution of the optimised non-stoquastic run and the stoquastic baseline: values under 1.0 (dotted line) indicate worse performance, while values over 1.0 are improved by the non-stoquastic catalyst. Our results suggest that the magnitude of the improvement is much better in the worst-case dataset, with some examples reaching 20x acceleration.}}
    \label{fig:non-stoquasticity-relative}
\end{figure}

\begin{figure}
    \centering
    \includegraphics[width=\linewidth]{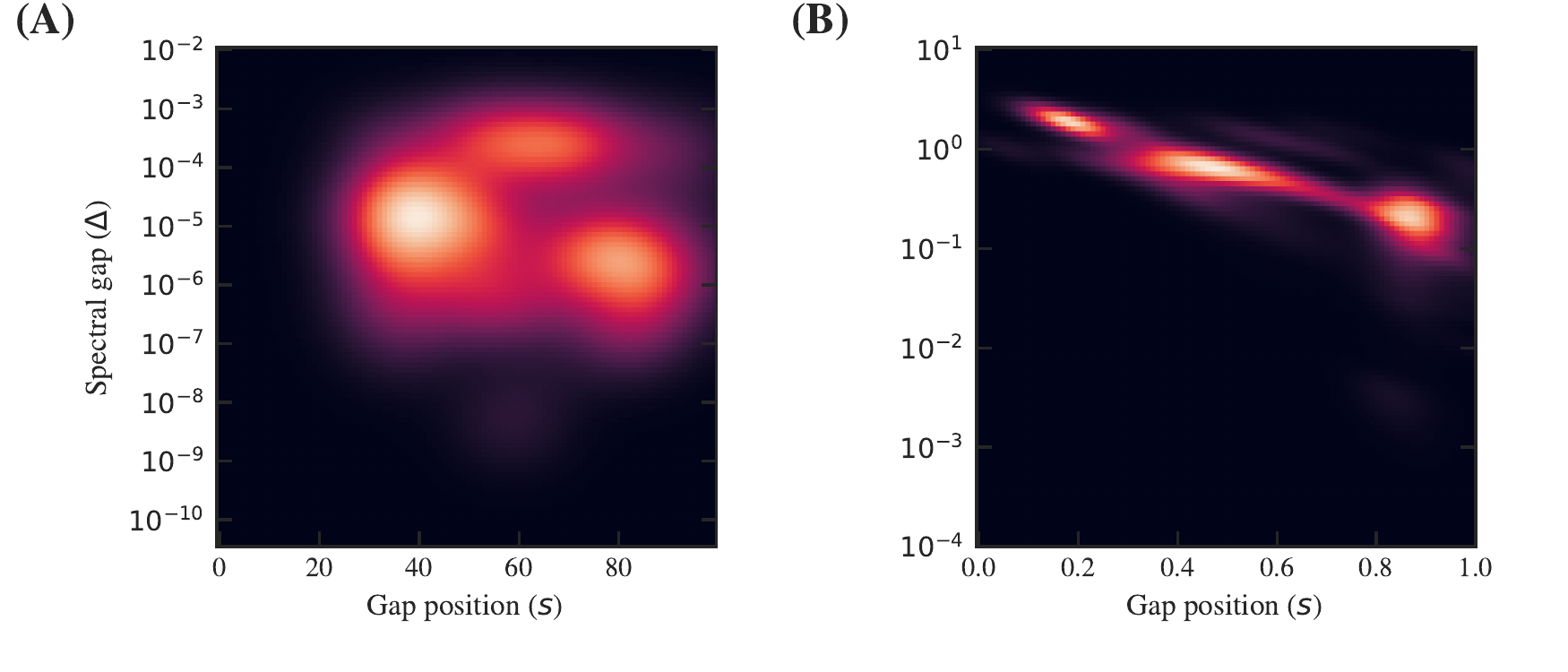}
    \caption{{ Distribution of spectral gap positions and magnitudes for 2D (A) and 3D (B) peptide sequences. The distributions have been estimated using Gaussian kernel density estimation (KDE). In the case of 2D peptides, we have removed the gaps greater than $10^{-3}$ to improve the visualization of smaller gaps; in the right plot we have used all the gaps in the dataset.}}
    \label{fig:gap-distribution}
\end{figure}

\begin{figure}
    \centering
    \includegraphics[width=\linewidth]{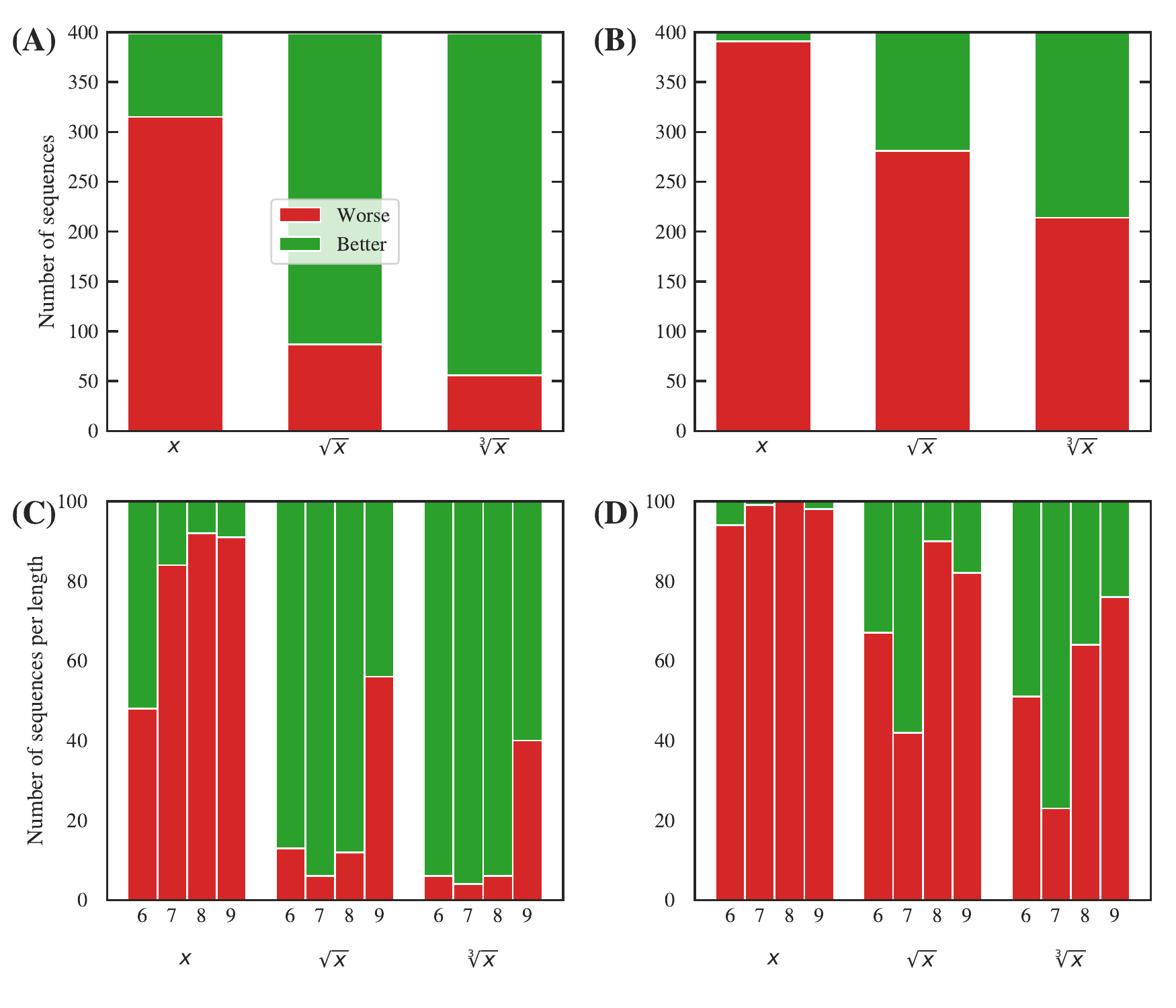}
    \caption{{ Variation of the annealing process with an annealing schedule built to slow down at regions most likely to display the minimal gap for any length. We consider linear ($x$), square root ($\sqrt{x}$) and cubic root ($\sqrt[3]{x}$) functions to compute the R-score (see Section \ref{section:methods}). (A, B) Proportion of solutions with a worse (red), identical (blue) or better (green) expected runtime than the baseline, for the worst-case (A) and random (B) datasets. (C, D) Proportions classified by length for the worst-case (C) and random (D) datasets. }}
    \label{fig:optimal-program}
\end{figure}

\begin{figure}
    \centering
    \includegraphics[width=\linewidth]{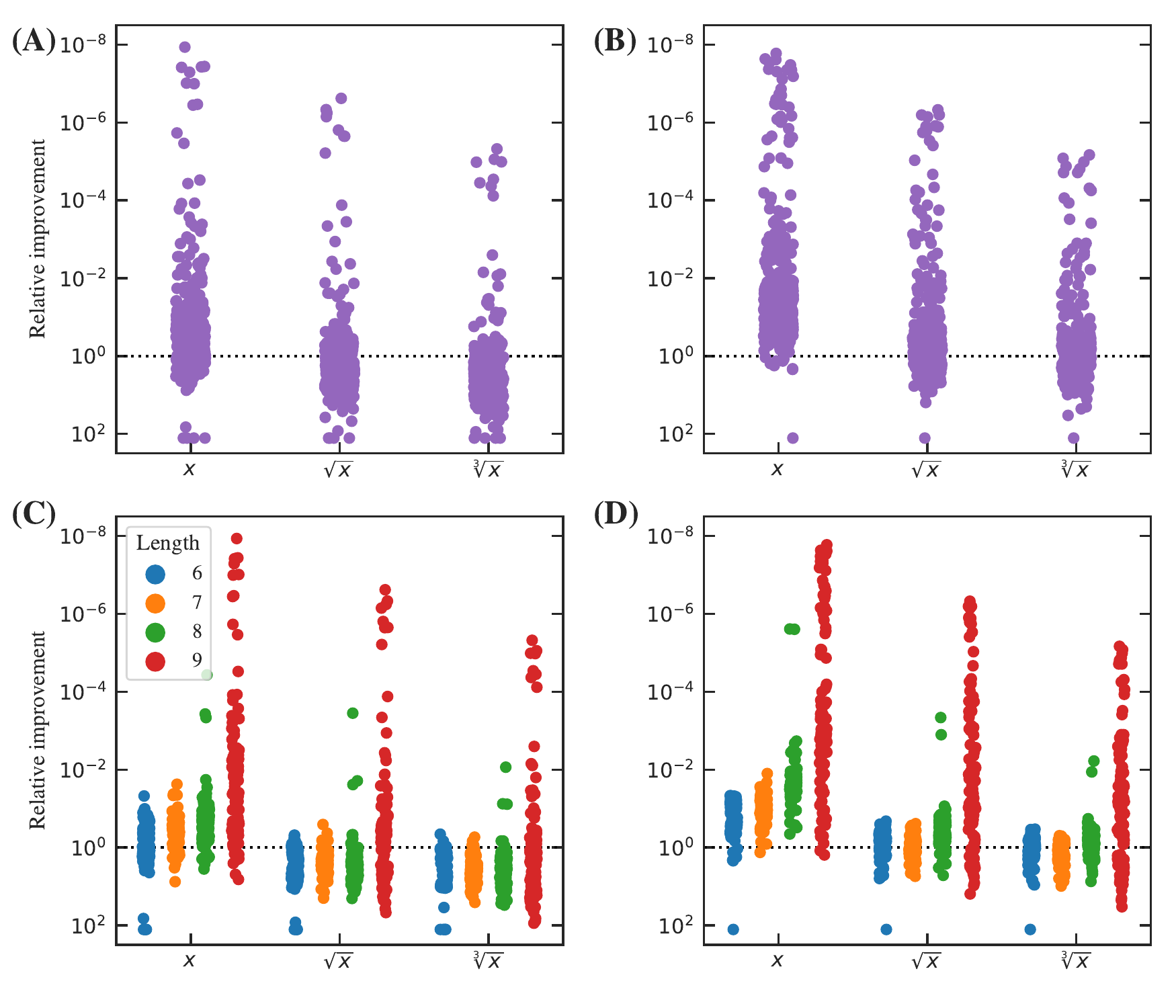}
    \caption{{ Relative improvement of the annealing process with an annealing schedule built to slow down at regions most likely to display the minimal gap for any length. We consider linear ($x$), square root ($\sqrt{x}$) and cubic root ($\sqrt[3]{x}$) functions to compute the R-score (see Section \ref{section:methods}). (A, B) Relative improvement of the expected time to solution time for the worst-case (A) and random (B) datasets. (C, D) Relative improvement classified by length for the worst-case (C) and random (D) datasets. Relative improvement here is defined as the ratio between the expected time to solution of  the  tailored trajectory over the  baseline:  values  under  1.0  (dotted  line) indicate worse performance, while values over 1.0 are improved by the tailored annealing schedule.}}
    \label{fig:optimal-program-relative}
\end{figure}

\begin{figure}
    \centering
    \includegraphics[width=\linewidth]{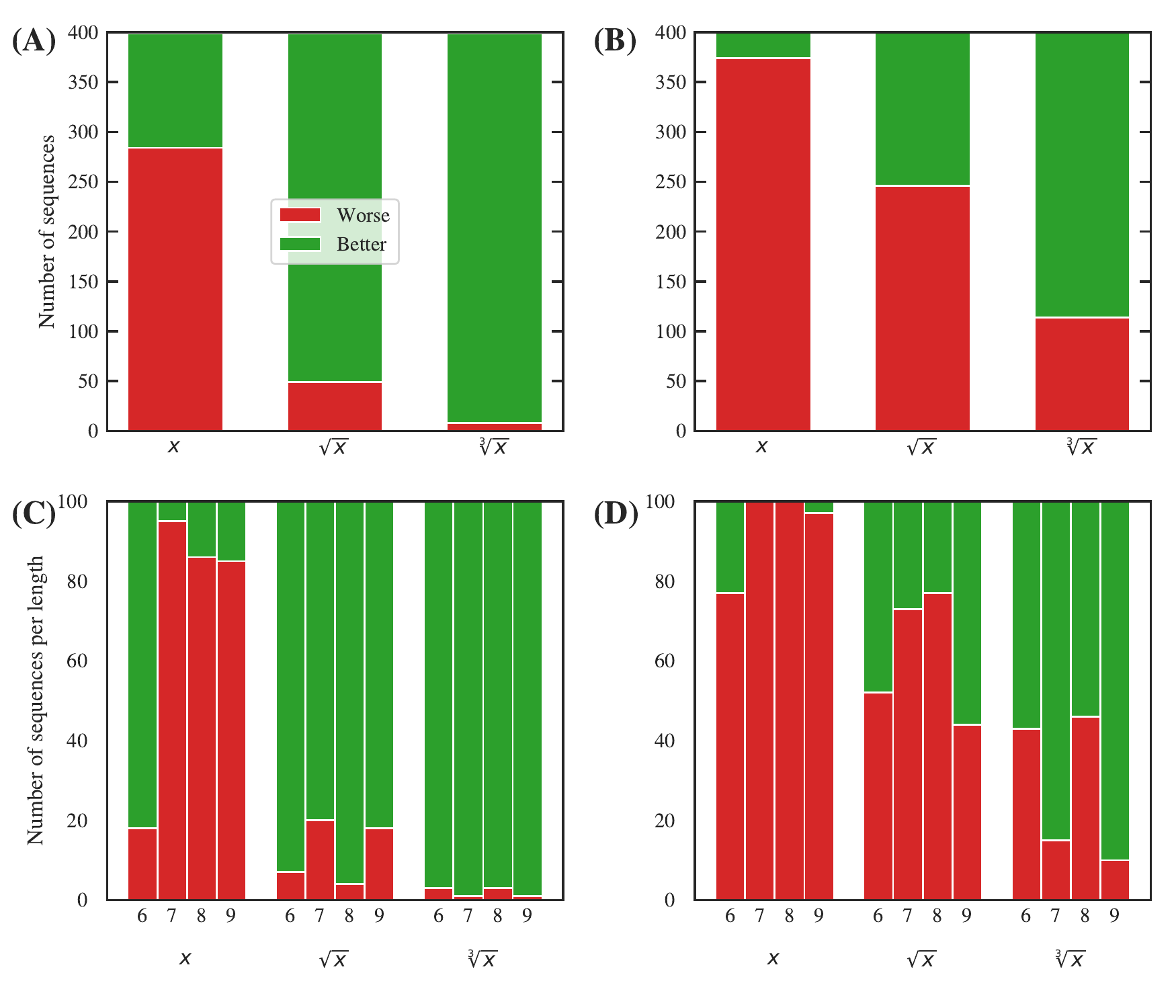}
    \caption{{ Variation of the annealing process with an annealing schedule built to slow down at regions most likely to display the minimal gap for the peptide's length. We consider linear ($x$), square root ($\sqrt{x}$) and cubic root ($\sqrt[3]{x}$) functions to compute the R-score (see Section \ref{section:methods}). (A, B) Proportion of solutions with a worse (red), identical (blue) or better (green) expected runtime than the baseline, for the worst-case (A) and random (B) datasets. (C, D) Proportions classified by length for the worst-case (C) and random (D) datasets. }}
    \label{fig:length-program}
\end{figure}

\begin{figure}
    \centering
    \includegraphics[width=\linewidth]{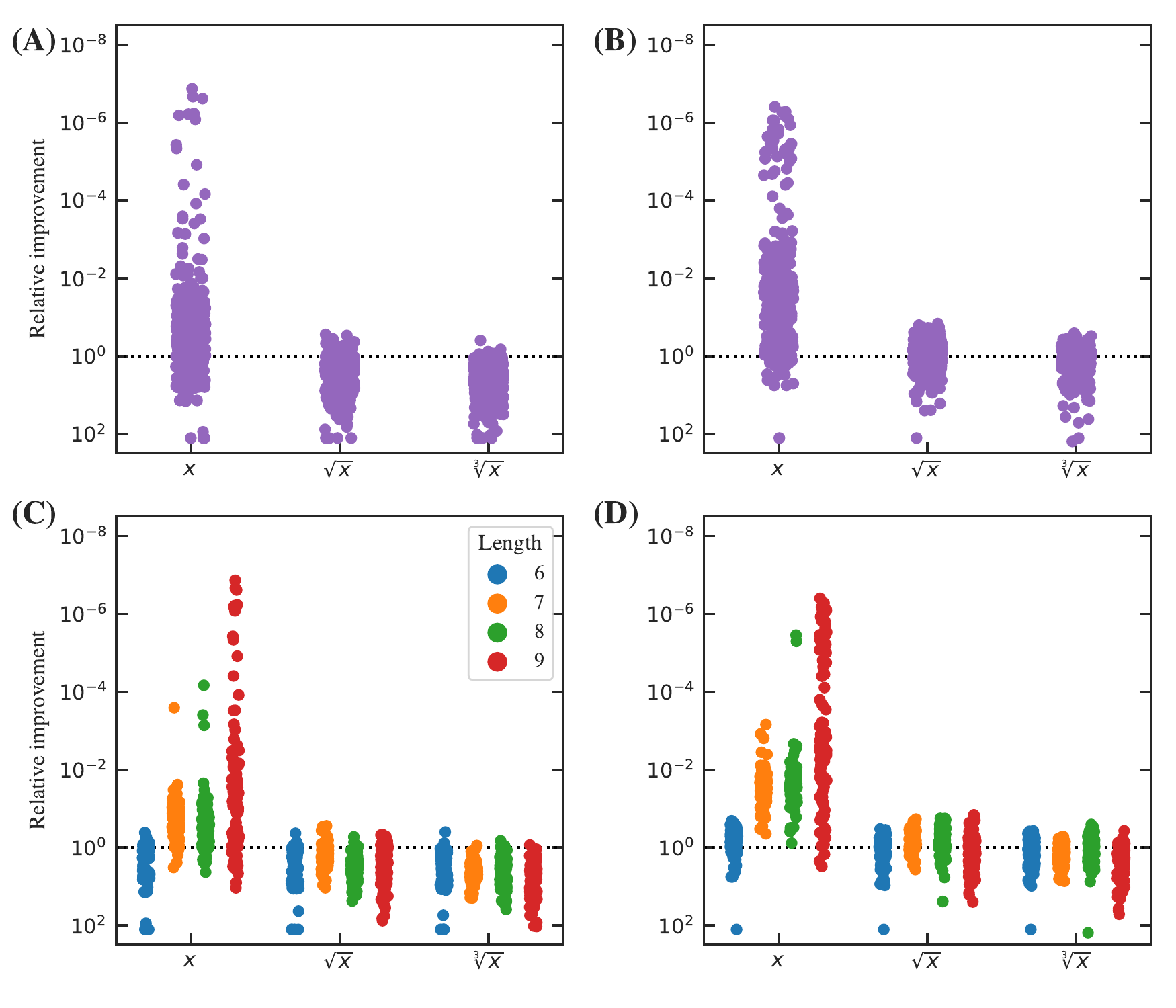}
    \caption{{ Relative improvement of the annealing process with an annealing schedule built to slow down at regions most likely to display the minimal gap for the peptide's length. We consider linear ($x$), square root ($\sqrt{x}$) and cubic root ($\sqrt[3]{x}$) functions to compute the R-score (see Section \ref{section:methods}). (A, B) Relative improvement of the expected time to solution time for the worst-case (A) and random (B) datasets. (C, D) Relative improvement classified by length for the worst-case (C) and random (D) datasets. Relative improvement here is defined as the ratio between the expected time to solution of  the  tailored trajectory over the  baseline:  values  under  1.0  (dotted  line) indicate worse performance, while values over 1.0 are improved by the tailored annealing schedule.}}
    \label{fig:length-program-relative}
\end{figure}

\FloatBarrier

\color{black}
\section{Model comparison raw data}
\label{appendix:statistics}

\begin{sidewaystable}
  \vspace{15cm}
  \centering
  \resizebox{\textwidth}{!}{%
  \begin{tabular}{ccccccccc}
    \toprule
    Parameter & Model  & 0-10\% & 10-25\% & 25-50\% & 50-75\% & 75-90\% & 90-100\% & Full \\
    \toprule
    $\alpha$ & $x^\alpha$ & $-3.286 \pm 0.053$ & $-1.605 \pm 0.012$ & $-0.779 \pm 0.005$ & $-0.184 \pm 0.005$ & $0.148 \pm 0.008$ & $0.361 \pm 0.013$ & $-0.752 \pm 0.010$ \\
    $\alpha$ & $e^{\alpha x}$ & $-1.598 \pm 0.028$ & $-0.781 \pm 0.007$ & $-0.379 \pm 0.003$ & $-0.088 \pm 0.002$ & $0.073 \pm 0.004$ & $0.173 \pm 0.007$ & $-0.366 \pm 0.005$ \\
    $\alpha$ & $e^{\alpha x^2}$ & $-0.552 \pm 0.012$ & $-0.269 \pm 0.003$ & $-0.130 \pm 0.001$ & $-0.029 \pm 0.001$ & $0.025 \pm 0.001$ & $0.057 \pm 0.003$ & $-0.126 \pm 0.002$ \\
    $\alpha$ & $e^{\alpha x^3}$ & $-0.181 \pm 0.005$ & $-0.088 \pm 0.001$ & $-0.042 \pm 0.001$ & $-0.009 \pm 0.000$ & $0.008 \pm 0.001$ & $0.018 \pm 0.001$ & $-0.041 \pm 0.001$ \\
    \midrule
    AIC & $x^\alpha$ & \textbf{2210.30} & \textbf{-2566.22} & \textbf{-9641.00} & \textbf{-9715.92} & \textbf{-4488.14} & \textbf{-2076.36} & \textbf{7421.17} \\
    AIC & $e^{\alpha x}$ & 2381.08 & -1849.16 & -8455.74 & -9587.50 & -4474.26 & -2016.89 & 7670.17 \\
    AIC & $e^{\alpha x^2}$ & 2771.33 & -540.84 & -6273.76 & -9297.51 & -4429.33 & -1884.47 & 8348.67 \\
    AIC & $e^{\alpha x^3}$ & 3021.55 & 159.84 & -5094.38 & -9098.27 & -4389.71 & -1794.93 & 8876.53 \\
    \midrule
    BIC & $x^\alpha$ & \textbf{2215.63} & \textbf{-2560.49} & \textbf{-9634.76} & \textbf{-9709.68} & \textbf{-4482.41} & \textbf{-2071.03} & \textbf{7428.80} \\
    BIC & $e^{\alpha x}$ & 2386.40 & -1843.43 & -8449.50 & -9581.26 & -4468.53 & -2011.56 & 7677.80 \\
    BIC & $e^{\alpha x^2}$ & 2776.65 & -535.11 & -6267.52 & -9291.27 & -4423.60 & -1879.14 & 8356.30 \\
    BIC & $e^{\alpha x^3}$ & 3026.87 & 165.57 & -5088.14 & -9092.03 & -4383.98 & -1789.60 & 8884.15 \\
    \midrule
    MSE & $x^\alpha$ & 22.43548 & \textbf{2.51534} & \textbf{0.20113} & \textbf{1.58745} & 3.47287 & 5.09703 & \textbf{0.20860} \\
    MSE & $e^{\alpha x}$ & 20.79187 & 2.51687 & 0.39273 & 1.67794 & 3.39960 & 4.83987 & 0.39967 \\
    MSE & $e^{\alpha x^2}$ & 18.27669 & 2.74269 & 0.86278 & 1.87008 & 3.23616 & 4.32081 & 0.86482 \\
    MSE & $e^{\alpha x^3}$ & \textbf{16.62559} & 2.85132 & 1.15869 & 1.99124 & \textbf{3.12596} & \textbf{3.99454} & 1.15846 \\
    \bottomrule
  \end{tabular}
  }
  \caption{Table of statistical fits for the minimum spectral gap between the ground state and the first excited state, for the dataset of 2D protein problems. In bold, we indicate the model preferred by a particular model selection criterion {(see Section \ref{section:methods})}.}
  \label{table:gap_all_2d}
\end{sidewaystable}

\begin{sidewaystable}
  \vspace{15cm}
  \centering
  \resizebox{\textwidth}{!}{%
  \begin{tabular}{ccccccccc}
    \toprule
    Parameter & Model  & 0-10\% & 10-25\% & 25-50\% & 50-75\% & 75-90\% & 90-100\% & Full \\
    \toprule
    $\alpha$ & $x^\alpha$ & $-3.665 \pm 0.051$ & $-1.575 \pm 0.011$ & $-0.850 \pm 0.004$ & $0.391 \pm 0.012$ & $1.162 \pm 0.010$ & $1.397 \pm 0.015$ & $-0.403 \pm 0.014$ \\
    $\alpha$ & $e^{\alpha x}$ & $-2.131 \pm 0.029$ & $-0.911 \pm 0.007$ & $-0.491 \pm 0.002$ & $0.239 \pm 0.007$ & $0.680 \pm 0.006$ & $0.798 \pm 0.010$ & $-0.231 \pm 0.008$ \\
    $\alpha$ & $e^{\alpha x^2}$ & $-1.121 \pm 0.017$ & $-0.473 \pm 0.005$ & $-0.254 \pm 0.002$ & $0.139 \pm 0.004$ & $0.363 \pm 0.003$ & $0.403 \pm 0.007$ & $-0.117 \pm 0.004$ \\
    $\alpha$ & $e^{\alpha x^3}$ & $-0.553 \pm 0.010$ & $-0.231 \pm 0.003$ & $-0.124 \pm 0.001$ & $0.072 \pm 0.002$ & $0.180 \pm 0.002$ & $0.194 \pm 0.004$ & $-0.056 \pm 0.002$ \\
    \midrule
    AIC & $x^\alpha$ & 1021.64 & \textbf{-4169.39} & \textbf{-12803.81} & -4379.94 & -4437.00 & \textbf{-2481.66} & \textbf{5915.36} \\
    AIC & $e^{\alpha x}$ & \textbf{992.01} & -4058.99 & -12412.42 & -4490.25 & \textbf{-4677.59} & -2250.96 & 5936.65 \\
    AIC & $e^{\alpha x^2}$ & 1249.22 & -2938.47 & -9875.84 & -4675.12 & -4325.99 & -1550.74 & 6033.55 \\
    AIC & $e^{\alpha x^3}$ & 1510.35 & -2213.93 & -8464.09 & \textbf{-4729.63} & -3775.90 & -1173.71 & 6110.65 \\
    \midrule
    BIC & $x^\alpha$ & 1026.91 & \textbf{-4163.71} & \textbf{-12797.62} & -4373.75 & -4431.33 & \textbf{-2476.39} & \textbf{5922.93} \\
    BIC & $e^{\alpha x}$ & \textbf{997.28} & -4053.32 & -12406.23 & -4484.06 & \textbf{-4671.92} & -2245.69 & 5944.22 \\
    BIC & $e^{\alpha x^2}$ & 1254.49 & -2932.80 & -9869.66 & -4668.93 & -4320.32 & -1545.48 & 6041.12 \\
    BIC & $e^{\alpha x^3}$ & 1515.61 & -2208.26 & -8457.91 & \textbf{-4723.44} & -3770.23 & -1168.44 & 6118.22 \\
    \midrule
    MSE & $x^\alpha$ & 17.94229 & 2.31403 & \textbf{0.33712} & \textbf{1.06890} & 4.14115 & 5.47780 & \textbf{0.00191} \\
    MSE & $e^{\alpha x}$ & 18.05314 & 2.31351 & 0.34474 & 1.11557 & 4.16563 & 5.31365 & 0.00912 \\
    MSE & $e^{\alpha x^2}$ & 17.15902 & 2.18661 & 0.35862 & 1.16065 & 3.96442 & 4.65149 & 0.04226 \\
    MSE & $e^{\alpha x^3}$ & \textbf{16.05752} & \textbf{2.05104} & 0.36410 & 1.14990 & \textbf{3.71657} & \textbf{4.14621} & 0.06887 \\
    \midrule
    \bottomrule
  \end{tabular}
  }
  \caption{Table of statistical fits for the minimum spectral gap between the ground state and the first excited state, for the dataset of 3D protein problems. In bold, we indicate the model preferred by a particular model selection criterion {(see Section \ref{section:methods})}.}
  \label{table:gap_all_3d}
\end{sidewaystable}
\begin{sidewaystable}
  \vspace{15cm}
  \centering
  \resizebox{\textwidth}{!}{%
  \begin{tabular}{ccccccccc}
    \toprule
    Parameter & Model  & 0-10\% & 10-25\% & 25-50\% & 50-75\% & 75-90\% & 90-100\% & Full \\
    \toprule
    $\alpha$ & $x^\alpha$ & $0.031 \pm 0.058$ & $0.265 \pm 0.026$ & $0.460 \pm 0.010$ & $0.795 \pm 0.015$ & $1.107 \pm 0.027$ & $1.510 \pm 0.060$ & $0.672 \pm 0.025$ \\
    $\alpha$ & $e^{\alpha x}$ & $0.009 \pm 0.013$ & $0.059 \pm 0.006$ & $0.101 \pm 0.002$ & $0.176 \pm 0.003$ & $0.245 \pm 0.005$ & $0.333 \pm 0.013$ & $0.149 \pm 0.005$ \\
    $\alpha$ & $e^{\alpha x^2}$ & $0.005 \pm 0.005$ & $0.022 \pm 0.002$ & $0.036 \pm 0.002$ & $0.064 \pm 0.002$ & $0.089 \pm 0.004$ & $0.120 \pm 0.008$ & $0.054 \pm 0.002$ \\
    $\alpha$ & $e^{\alpha x^3}$ & $0.002 \pm 0.002$ & $0.007 \pm 0.001$ & $0.012 \pm 0.001$ & $0.021 \pm 0.001$ & $0.030 \pm 0.002$ & $0.040 \pm 0.003$ & $0.018 \pm 0.001$ \\
    \midrule
    AIC & $x^\alpha$ & -149.70 & -287.88 & \textbf{-612.57} & -527.89 & -276.94 & \textbf{-147.00} & -1240.01 \\
    AIC & $e^{\alpha x}$ & -149.96 & \textbf{-288.86} & -592.28 & \textbf{-533.08} & \textbf{-287.01} & -145.83 & \textbf{-1243.19} \\
    AIC & $e^{\alpha x^2}$ & -150.45 & -281.14 & -491.06 & -409.62 & -213.34 & -112.17 & -1186.99 \\
    AIC & $e^{\alpha x^3}$ & \textbf{-150.55} & -272.31 & -442.33 & -352.09 & -178.27 & -93.24 & -1129.40 \\
    \midrule
    BIC & $x^\alpha$ & -148.01 & -285.82 & \textbf{-610.01} & -525.35 & -274.91 & \textbf{-145.31} & -1236.06 \\
    BIC & $e^{\alpha x}$ & -148.27 & \textbf{-286.80} & -589.71 & \textbf{-530.54} & \textbf{-284.98} & -144.14 & \textbf{-1239.24} \\
    BIC & $e^{\alpha x^2}$ & -148.76 & -279.08 & -488.49 & -407.08 & -211.31 & -110.48 & -1183.04 \\
    BIC & $e^{\alpha x^3}$ & \textbf{-148.86} & -270.25 & -439.76 & -349.55 & -176.24 & -91.55 & -1125.45 \\
    \midrule
    MSE & $x^\alpha$ & 0.28234 & 0.11476 & 0.03290 & 0.01252 & 0.13092 & 0.48034 & 0.00226 \\
    MSE & $e^{\alpha x}$ & 0.27206 & \textbf{0.11299} & \textbf{0.03275} & \textbf{0.01143} & \textbf{0.13057} & 0.47586 & \textbf{0.00097} \\
    MSE & $e^{\alpha x^2}$ & \textbf{0.26061} & 0.12761 & 0.05664 & 0.03512 & 0.14624 & 0.45802 & 0.02540 \\
    MSE & $e^{\alpha x^3}$ & 0.26128 & 0.14645 & 0.08373 & 0.06295 & 0.16469 & \textbf{0.44568} & 0.05436 \\
    \midrule
    \bottomrule
  \end{tabular}
  }
  \caption{Table of statistical fits for the expected quantum annealing runtime for the worst-case dataset in 2D. In bold, we indicate the model preferred by a particular model selection criterion {(see Section \ref{section:methods})}.}
  \label{table:time_bad_2d}
\end{sidewaystable}
\begin{sidewaystable}
  \vspace{15cm}
  \centering
  \resizebox{\textwidth}{!}{%
  \begin{tabular}{ccccccccc}
    \toprule
    Parameter & Model  & 0-10\% & 10-25\% & 25-50\% & 50-75\% & 75-90\% & 90-100\% & Full \\
    \toprule
    $\alpha$ & $x^\alpha$ & $0.197 \pm 0.034$ & $0.313 \pm 0.025$ & $0.497 \pm 0.017$ & $0.728 \pm 0.015$ & $0.928 \pm 0.033$ & $1.294 \pm 0.065$ & $0.642 \pm 0.020$ \\
    $\alpha$ & $e^{\alpha x}$ & $0.046 \pm 0.007$ & $0.071 \pm 0.005$ & $0.111 \pm 0.004$ & $0.161 \pm 0.004$ & $0.203 \pm 0.009$ & $0.282 \pm 0.017$ & $0.142 \pm 0.004$ \\
    $\alpha$ & $e^{\alpha x^2}$ & $0.018 \pm 0.003$ & $0.027 \pm 0.002$ & $0.041 \pm 0.002$ & $0.058 \pm 0.002$ & $0.073 \pm 0.005$ & $0.099 \pm 0.009$ & $0.052 \pm 0.002$ \\
    $\alpha$ & $e^{\alpha x^3}$ & $0.006 \pm 0.001$ & $0.009 \pm 0.001$ & $0.014 \pm 0.001$ & $0.020 \pm 0.001$ & $0.024 \pm 0.002$ & $0.033 \pm 0.004$ & $0.017 \pm 0.001$ \\
    \midrule
    AIC & $x^\alpha$ & -186.48 & -282.65 & -491.03 & \textbf{-514.65} & \textbf{-248.51} & \textbf{-137.01} & \textbf{-1357.19} \\
    AIC & $e^{\alpha x}$ & -190.18 & \textbf{-290.24} & \textbf{-499.15} & -501.84 & -231.11 & -125.93 & -1351.74 \\
    AIC & $e^{\alpha x^2}$ & \textbf{-193.61} & -289.73 & -456.11 & -402.68 & -186.83 & -98.81 & -1274.09 \\
    AIC & $e^{\alpha x^3}$ & -192.53 & -281.21 & -424.12 & -358.44 & -166.78 & -86.11 & -1212.57 \\
    \midrule
    BIC & $x^\alpha$ & -184.84 & -280.66 & -488.53 & \textbf{-512.15} & \textbf{-246.54} & \textbf{-135.37} & \textbf{-1353.30} \\
    BIC & $e^{\alpha x}$ & -188.54 & \textbf{-288.25} & \textbf{-496.65} & -499.34 & -229.14 & -124.29 & -1347.85 \\
    BIC & $e^{\alpha x^2}$ & \textbf{-191.97} & -287.74 & -453.61 & -400.18 & -184.86 & -97.17 & -1270.20 \\
    BIC & $e^{\alpha x^3}$ & -190.89 & -279.22 & -421.62 & -355.94 & -164.81 & -84.47 & -1208.68 \\
    \midrule
    MSE & $x^\alpha$ & 0.14249 & 0.08153 & \textbf{0.02162} & \textbf{0.01197} & 0.06227 & 0.29482 & \textbf{0.00706} \\
    MSE & $e^{\alpha x}$ & \textbf{0.13655} & \textbf{0.07829} & 0.02171 & 0.01340 & \textbf{0.06107} & 0.28277 & 0.00824 \\
    MSE & $e^{\alpha x^2}$ & 0.13927 & 0.08846 & 0.04172 & 0.03587 & 0.07512 & 0.25662 & 0.03091 \\
    MSE & $e^{\alpha x^3}$ & 0.14835 & 0.10230 & 0.06192 & 0.05742 & 0.09064 & \textbf{0.24276} & 0.05290 \\
    \midrule
    \bottomrule
  \end{tabular}
  }
  \caption{Table of statistical fits for the expected quantum annealing runtime for the random dataset in 2D. In bold, we indicate the model preferred by a particular model selection criterion {(see Section \ref{section:methods})}.}
  \label{table:time_random_2d}
\end{sidewaystable}
\begin{sidewaystable}
  \vspace{15cm}
  \centering
  \resizebox{\textwidth}{!}{%
  \begin{tabular}{ccccccccc}
    \toprule
    Parameter & Model  & 0-10\% & 10-25\% & 25-50\% & 50-75\% & 75-90\% & 90-100\% & Full \\
    \toprule
    $\alpha$ & $x^\alpha$ & $1.241 \pm 0.137$ & $1.344 \pm 0.096$ & $1.491 \pm 0.055$ & $1.773 \pm 0.021$ & $2.002 \pm 0.053$ & $2.343 \pm 0.171$ & $1.676 \pm 0.036$ \\
    $\alpha$ & $e^{\alpha x}$ & $0.329 \pm 0.029$ & $0.351 \pm 0.019$ & $0.384 \pm 0.010$ & $0.445 \pm 0.006$ & $0.496 \pm 0.019$ & $0.571 \pm 0.052$ & $0.424 \pm 0.009$ \\
    $\alpha$ & $e^{\alpha x^2}$ & $0.191 \pm 0.010$ & $0.198 \pm 0.006$ & $0.211 \pm 0.003$ & $0.231 \pm 0.008$ & $0.250 \pm 0.018$ & $0.274 \pm 0.039$ & $0.224 \pm 0.006$ \\
    $\alpha$ & $e^{\alpha x^3}$ & $0.099 \pm 0.004$ & $0.102 \pm 0.002$ & $0.107 \pm 0.002$ & $0.113 \pm 0.006$ & $0.120 \pm 0.011$ & $0.127 \pm 0.023$ & $0.111 \pm 0.003$ \\
    \midrule
    AIC & $x^\alpha$ & -83.43 & -137.51 & -276.34 & \textbf{-412.16} & \textbf{-189.25} & \textbf{-70.28} & -937.74 \\
    AIC & $e^{\alpha x}$ & -93.59 & -156.59 & -324.99 & -391.41 & -158.26 & -59.27 & \textbf{-961.33} \\
    AIC & $e^{\alpha x^2}$ & -122.89 & -212.15 & \textbf{-408.62} & -258.19 & -109.09 & -40.07 & -874.59 \\
    AIC & $e^{\alpha x^3}$ & \textbf{-138.75} & \textbf{-227.55} & -343.24 & -213.18 & -90.80 & -32.14 & -785.15 \\
    \midrule
    BIC & $x^\alpha$ & -82.03 & -135.73 & -274.04 & \textbf{-409.87} & \textbf{-187.46} & \textbf{-68.88} & -934.05 \\
    BIC & $e^{\alpha x}$ & -92.19 & -154.81 & -322.69 & -389.12 & -156.47 & -57.87 & \textbf{-957.64} \\
    BIC & $e^{\alpha x^2}$ & -121.49 & -210.37 & \textbf{-406.31} & -255.90 & -107.31 & -38.67 & -870.91 \\
    BIC & $e^{\alpha x^3}$ & \textbf{-137.35} & \textbf{-225.77} & -340.93 & -210.89 & -89.02 & -30.74 & -781.46 \\
    \midrule
    MSE & $x^\alpha$ & 0.06899 & 0.04407 & 0.02023 & 0.01300 & 0.04434 & 0.15289 & 0.00975 \\
    MSE & $e^{\alpha x}$ & \textbf{0.04570} & \textbf{0.02694} & \textbf{0.00800} & \textbf{0.00221} & \textbf{0.02562} & 0.10706 & \textbf{0.00002} \\
    MSE & $e^{\alpha x^2}$ & 0.06041 & 0.05259 & 0.04358 & 0.04114 & 0.05080 & \textbf{0.08194} & 0.04043 \\
    MSE & $e^{\alpha x^3}$ & 0.10648 & 0.10266 & 0.09807 & 0.09714 & 0.10178 & 0.11401 & 0.09681 \\
    \midrule
    \bottomrule
  \end{tabular}
  }
  \caption{Table of statistical fits for the expected quantum annealing runtime for the worst-case dataset in 3D. In bold, we indicate the model preferred by a particular model selection criterion {(see Section \ref{section:methods})}.}
  \label{table:time_bad_3d}
\end{sidewaystable}
\begin{sidewaystable}
  \vspace{15cm}
  \centering
  \resizebox{\textwidth}{!}{%
  \begin{tabular}{ccccccccc}
    \toprule
    Parameter & Model  & 0-10\% & 10-25\% & 25-50\% & 50-75\% & 75-90\% & 90-100\% & Full \\
    \toprule
    $\alpha$ & $x^\alpha$ & $1.644 \pm 0.122$ & $1.733 \pm 0.091$ & $1.816 \pm 0.060$ & $1.942 \pm 0.054$ & $2.200 \pm 0.035$ & $2.815 \pm 0.175$ & $1.978 \pm 0.038$ \\
    $\alpha$ & $e^{\alpha x}$ & $0.429 \pm 0.023$ & $0.448 \pm 0.016$ & $0.469 \pm 0.010$ & $0.499 \pm 0.008$ & $0.554 \pm 0.007$ & $0.688 \pm 0.055$ & $0.504 \pm 0.008$ \\
    $\alpha$ & $e^{\alpha x^2}$ & $0.241 \pm 0.004$ & $0.248 \pm 0.002$ & $0.258 \pm 0.002$ & $0.272 \pm 0.003$ & $0.290 \pm 0.012$ & $0.334 \pm 0.043$ & $0.271 \pm 0.005$ \\
    $\alpha$ & $e^{\alpha x^3}$ & $0.123 \pm 0.002$ & $0.126 \pm 0.002$ & $0.130 \pm 0.003$ & $0.136 \pm 0.004$ & $0.143 \pm 0.009$ & $0.156 \pm 0.025$ & $0.135 \pm 0.003$ \\
    \midrule
    AIC & $x^\alpha$ & -90.49 & -142.04 & -256.61 & -272.30 & -221.56 & \textbf{-69.03} & -903.10 \\
    AIC & $e^{\alpha x}$ & -107.31 & -170.90 & -319.80 & -349.58 & \textbf{-236.72} & -55.79 & \textbf{-975.92} \\
    AIC & $e^{\alpha x^2}$ & \textbf{-170.86} & \textbf{-283.27} & \textbf{-491.91} & \textbf{-395.21} & -142.57 & -33.77 & -920.09 \\
    AIC & $e^{\alpha x^3}$ & -166.46 & -230.03 & -325.45 & -283.26 & -112.39 & -24.94 & -803.86 \\
    \midrule
    BIC & $x^\alpha$ & -89.09 & -140.26 & -254.34 & -270.02 & -219.80 & \textbf{-67.63} & -899.42 \\
    BIC & $e^{\alpha x}$ & -105.91 & -169.11 & -317.52 & -347.30 & \textbf{-234.96} & -54.39 & \textbf{-972.24} \\
    BIC & $e^{\alpha x^2}$ & \textbf{-169.46} & \textbf{-281.48} & \textbf{-489.63} & \textbf{-392.93} & -140.81 & -32.37 & -916.41 \\
    BIC & $e^{\alpha x^3}$ & -165.05 & -228.24 & -323.18 & -280.98 & -110.63 & -23.54 & -800.19 \\
    \midrule
    MSE & $x^\alpha$ & 0.07051 & 0.05430 & 0.04381 & 0.03617 & 0.05208 & 0.26086 & 0.03583 \\
    MSE & $e^{\alpha x}$ & \textbf{0.03439} & \textbf{0.02162} & \textbf{0.01238} & \textbf{0.00623} & \textbf{0.01876} & 0.17519 & \textbf{0.00609} \\
    MSE & $e^{\alpha x^2}$ & 0.04346 & 0.03699 & 0.03132 & 0.02834 & 0.03476 & \textbf{0.09525} & 0.02834 \\
    MSE & $e^{\alpha x^3}$ & 0.10002 & 0.09602 & 0.09251 & 0.09105 & 0.09499 & 0.11941 & 0.09093 \\
    \midrule
    \bottomrule
  \end{tabular}
  }
  \caption{Table of statistical fits for the expected quantum annealing runtime for the random dataset in 3D. In bold, we indicate the model preferred by a particular model selection criterion {(see Section \ref{section:methods})}.}
  \label{table:time_random_3d}
\end{sidewaystable}
\begin{sidewaystable}
  \vspace{15cm}
  \centering
  \resizebox{\textwidth}{!}{%
  \begin{tabular}{ccccccccc}
    \toprule
    Parameter & Model  & 0-10\% & 10-25\% & 25-50\% & 50-75\% & 75-90\% & 90-100\% & Full \\
    \toprule
    $\alpha$ & $x^\alpha$ & $0.342 \pm 0.024$ & $0.370 \pm 0.022$ & $0.406 \pm 0.017$ & $0.441 \pm 0.019$ & $0.474 \pm 0.027$ & $0.582 \pm 0.043$ & $0.431 \pm 0.010$ \\
    $\alpha$ & $e^{\alpha x}$ & $0.079 \pm 0.004$ & $0.086 \pm 0.003$ & $0.094 \pm 0.003$ & $0.102 \pm 0.003$ & $0.110 \pm 0.004$ & $0.134 \pm 0.007$ & $0.100 \pm 0.002$ \\
    $\alpha$ & $e^{\alpha x^2}$ & $0.031 \pm 0.000$ & $0.034 \pm 0.000$ & $0.037 \pm 0.000$ & $0.040 \pm 0.000$ & $0.043 \pm 0.000$ & $0.053 \pm 0.001$ & $0.039 \pm 0.000$ \\
    $\alpha$ & $e^{\alpha x^3}$ & $0.011 \pm 0.000$ & $0.012 \pm 0.000$ & $0.013 \pm 0.000$ & $0.014 \pm 0.000$ & $0.015 \pm 0.000$ & $0.019 \pm 0.000$ & $0.014 \pm 0.000$ \\
    \midrule
    AIC & $x^\alpha$ & -221.09 & -302.58 & -502.23 & -484.59 & -281.00 & -174.66 & -1909.64 \\
    AIC & $e^{\alpha x}$ & -249.38 & -343.60 & -575.55 & -558.84 & -325.25 & -199.98 & -2139.47 \\
    AIC & $e^{\alpha x^2}$ & \textbf{-368.49} & \textbf{-553.71} & \textbf{-956.21} & \textbf{-866.85} & \textbf{-525.90} & \textbf{-270.71} & \textbf{-2608.48} \\
    AIC & $e^{\alpha x^3}$ & -320.65 & -436.13 & -692.79 & -647.66 & -380.99 & -253.89 & -2428.29 \\
    \midrule
    BIC & $x^\alpha$ & -219.40 & -300.56 & -499.68 & -482.06 & -278.97 & -172.97 & -1905.70 \\
    BIC & $e^{\alpha x}$ & -247.69 & -341.57 & -573.00 & -556.31 & -323.22 & -198.29 & -2135.53 \\
    BIC & $e^{\alpha x^2}$ & \textbf{-366.80} & \textbf{-551.68} & \textbf{-953.67} & \textbf{-864.31} & \textbf{-523.87} & \textbf{-269.02} & \textbf{-2604.55} \\
    BIC & $e^{\alpha x^3}$ & -318.96 & -434.11 & -690.25 & -645.13 & -378.96 & -252.20 & -2424.35 \\
    \midrule
    MSE & $x^\alpha$ & 0.02763 & 0.02468 & 0.02233 & 0.02179 & 0.02282 & 0.03639 & 0.02177 \\
    MSE & $e^{\alpha x}$ & 0.01633 & 0.01286 & 0.01063 & 0.01007 & 0.01142 & 0.02644 & 0.01005 \\
    MSE & $e^{\alpha x^2}$ & \textbf{0.00662} & \textbf{0.00264} & \textbf{0.00075} & \textbf{0.00016} & \textbf{0.00190} & \textbf{0.01857} & \textbf{0.00013} \\
    MSE & $e^{\alpha x^3}$ & 0.00883 & 0.00492 & 0.00325 & 0.00266 & 0.00440 & 0.02136 & 0.00264 \\
    \midrule
    \bottomrule
  \end{tabular}
  }
  \caption{Table of statistical fits for the expected number of energy evaluations in classical simulated annealing for the worst-case dataset in 2D. In bold, we indicate the model preferred by a particular model selection criterion {(see Section \ref{section:methods})}.}
  \label{table:sa_bad_2d}
\end{sidewaystable}
\begin{sidewaystable}
  \vspace{15cm}
  \centering
  \resizebox{\textwidth}{!}{%
  \begin{tabular}{ccccccccc}
    \toprule
    Parameter & Model  & 0-10\% & 10-25\% & 25-50\% & 50-75\% & 75-90\% & 90-100\% & Full \\
    \toprule
    $\alpha$ & $x^\alpha$ & $0.291 \pm 0.022$ & $0.330 \pm 0.020$ & $0.366 \pm 0.017$ & $0.404 \pm 0.019$ & $0.450 \pm 0.027$ & $0.585 \pm 0.046$ & $0.397 \pm 0.010$ \\
    $\alpha$ & $e^{\alpha x}$ & $0.068 \pm 0.003$ & $0.077 \pm 0.003$ & $0.085 \pm 0.003$ & $0.094 \pm 0.003$ & $0.105 \pm 0.004$ & $0.136 \pm 0.008$ & $0.092 \pm 0.002$ \\
    $\alpha$ & $e^{\alpha x^2}$ & $0.027 \pm 0.000$ & $0.030 \pm 0.000$ & $0.034 \pm 0.000$ & $0.037 \pm 0.000$ & $0.041 \pm 0.000$ & $0.054 \pm 0.001$ & $0.036 \pm 0.000$ \\
    $\alpha$ & $e^{\alpha x^3}$ & $0.009 \pm 0.000$ & $0.011 \pm 0.000$ & $0.012 \pm 0.000$ & $0.013 \pm 0.000$ & $0.014 \pm 0.000$ & $0.019 \pm 0.001$ & $0.013 \pm 0.000$ \\
    \midrule
    AIC & $x^\alpha$ & -220.91 & -308.49 & -481.97 & -467.88 & -261.99 & -163.71 & -1819.85 \\
    AIC & $e^{\alpha x}$ & -247.33 & -347.44 & -544.89 & -532.60 & -299.81 & -186.41 & -2004.83 \\
    AIC & $e^{\alpha x^2}$ & \textbf{-349.17} & \textbf{-530.83} & \textbf{-832.72} & \textbf{-839.80} & \textbf{-481.49} & \textbf{-239.03} & \textbf{-2348.93} \\
    AIC & $e^{\alpha x^3}$ & -312.87 & -457.07 & -720.89 & -688.38 & -383.29 & -225.58 & -2278.73 \\
    \midrule
    BIC & $x^\alpha$ & -219.27 & -306.49 & -479.48 & -465.39 & -260.04 & -162.07 & -1815.96 \\
    BIC & $e^{\alpha x}$ & -245.69 & -345.43 & -542.41 & -530.11 & -297.86 & -184.77 & -2000.94 \\
    BIC & $e^{\alpha x^2}$ & \textbf{-347.53} & \textbf{-528.83} & \textbf{-830.23} & \textbf{-837.31} & \textbf{-479.54} & \textbf{-237.39} & \textbf{-2345.04} \\
    BIC & $e^{\alpha x^3}$ & -311.24 & -455.06 & -718.40 & -685.89 & -381.34 & -223.94 & -2274.85 \\
    \midrule
    MSE & $x^\alpha$ & 0.02863 & 0.02382 & 0.02123 & 0.02035 & 0.02197 & 0.04334 & 0.02036 \\
    MSE & $e^{\alpha x}$ & 0.01860 & 0.01347 & 0.01070 & 0.00984 & 0.01178 & 0.03607 & 0.00983 \\
    MSE & $e^{\alpha x^2}$ & \textbf{0.00937} & \textbf{0.00395} & \textbf{0.00104} & \textbf{0.00024} & \textbf{0.00262} & \textbf{0.02961} & \textbf{0.00022} \\
    MSE & $e^{\alpha x^3}$ & 0.01056 & 0.00520 & 0.00234 & 0.00158 & 0.00403 & 0.03065 & 0.00156 \\
    \midrule
    \bottomrule
  \end{tabular}
  }
  \caption{Table of statistical fits for the expected number of energy evaluations in classical simulated annealing for the random dataset in 2D. In bold, we indicate the model preferred by a particular model selection criterion {(see Section \ref{section:methods})}.}
  \label{table:sa_random_2d}
\end{sidewaystable}
\begin{sidewaystable}
  \vspace{15cm}
  \centering
  \resizebox{\textwidth}{!}{%
  \begin{tabular}{ccccccccc}
    \toprule
    Parameter & Model  & 0-10\% & 10-25\% & 25-50\% & 50-75\% & 75-90\% & 90-100\% & Full \\
    \toprule
    $\alpha$ & $x^\alpha$ & $0.331 \pm 0.024$ & $0.359 \pm 0.020$ & $0.395 \pm 0.017$ & $0.439 \pm 0.019$ & $0.464 \pm 0.026$ & $0.493 \pm 0.032$ & $0.414 \pm 0.009$ \\
    $\alpha$ & $e^{\alpha x}$ & $0.086 \pm 0.004$ & $0.093 \pm 0.004$ & $0.103 \pm 0.003$ & $0.114 \pm 0.004$ & $0.121 \pm 0.005$ & $0.128 \pm 0.006$ & $0.108 \pm 0.002$ \\
    $\alpha$ & $e^{\alpha x^2}$ & $0.048 \pm 0.001$ & $0.052 \pm 0.000$ & $0.057 \pm 0.000$ & $0.064 \pm 0.000$ & $0.068 \pm 0.000$ & $0.071 \pm 0.000$ & $0.060 \pm 0.000$ \\
    $\alpha$ & $e^{\alpha x^3}$ & $0.025 \pm 0.000$ & $0.027 \pm 0.000$ & $0.029 \pm 0.000$ & $0.032 \pm 0.000$ & $0.034 \pm 0.000$ & $0.036 \pm 0.001$ & $0.031 \pm 0.000$ \\
    \midrule
    AIC & $x^\alpha$ & -189.03 & -266.65 & -437.52 & -414.11 & -250.89 & -164.34 & -1697.35 \\
    AIC & $e^{\alpha x}$ & -207.15 & -294.11 & -483.51 & -459.26 & -278.90 & -184.10 & -1856.48 \\
    AIC & $e^{\alpha x^2}$ & \textbf{-295.05} & \textbf{-454.12} & \textbf{-717.37} & \textbf{-714.97} & \textbf{-444.30} & \textbf{-310.68} & \textbf{-2316.48} \\
    AIC & $e^{\alpha x^3}$ & -276.28 & -384.31 & -622.73 & -613.55 & -373.74 & -230.00 & -2217.73 \\
    \midrule
    BIC & $x^\alpha$ & -187.63 & -264.89 & -435.24 & -411.84 & -249.11 & -162.98 & -1693.68 \\
    BIC & $e^{\alpha x}$ & -205.74 & -292.35 & -481.23 & -457.00 & -277.12 & -182.73 & -1852.81 \\
    BIC & $e^{\alpha x^2}$ & \textbf{-293.65} & \textbf{-452.36} & \textbf{-715.09} & \textbf{-712.71} & \textbf{-442.52} & \textbf{-309.31} & \textbf{-2312.81} \\
    BIC & $e^{\alpha x^3}$ & -274.88 & -382.54 & -620.45 & -611.29 & -371.95 & -228.64 & -2214.06 \\
    \midrule
    MSE & $x^\alpha$ & 0.00943 & 0.00824 & 0.00740 & 0.00752 & 0.00813 & 0.00933 & 0.00730 \\
    MSE & $e^{\alpha x}$ & 0.00607 & 0.00484 & 0.00396 & 0.00405 & 0.00473 & 0.00595 & 0.00384 \\
    MSE & $e^{\alpha x^2}$ & \textbf{0.00239} & \textbf{0.00113} & \textbf{0.00022} & \textbf{0.00027} & \textbf{0.00102} & \textbf{0.00221} & \textbf{0.00007} \\
    MSE & $e^{\alpha x^3}$ & 0.00274 & 0.00152 & 0.00062 & 0.00066 & 0.00142 & 0.00253 & 0.00046 \\
    \midrule
    \bottomrule
  \end{tabular}
  }
  \caption{Table of statistical fits for the expected number of energy evaluations in classical simulated annealing for the worst-case dataset in 3D. In bold, we indicate the model preferred by a particular model selection criterion {(see Section \ref{section:methods})}.}
  \label{table:sa_bad_3d}
\end{sidewaystable}
\begin{sidewaystable}
  \vspace{15cm}
  \centering
    \resizebox{\textwidth}{!}{%
  \begin{tabular}{ccccccccc}
    \toprule
    Parameter & Model  & 0-10\% & 10-25\% & 25-50\% & 50-75\% & 75-90\% & 90-100\% & Full \\
    \toprule
    $\alpha$ & $x^\alpha$ & $0.307 \pm 0.022$ & $0.350 \pm 0.020$ & $0.393 \pm 0.018$ & $0.434 \pm 0.018$ & $0.471 \pm 0.025$ & $0.511 \pm 0.031$ & $0.412 \pm 0.009$ \\
    $\alpha$ & $e^{\alpha x}$ & $0.080 \pm 0.004$ & $0.091 \pm 0.004$ & $0.102 \pm 0.003$ & $0.113 \pm 0.003$ & $0.122 \pm 0.005$ & $0.132 \pm 0.005$ & $0.107 \pm 0.002$ \\
    $\alpha$ & $e^{\alpha x^2}$ & $0.045 \pm 0.001$ & $0.051 \pm 0.000$ & $0.057 \pm 0.000$ & $0.063 \pm 0.000$ & $0.068 \pm 0.000$ & $0.074 \pm 0.000$ & $0.060 \pm 0.001$ \\
    $\alpha$ & $e^{\alpha x^3}$ & $0.023 \pm 0.000$ & $0.026 \pm 0.000$ & $0.029 \pm 0.000$ & $0.032 \pm 0.000$ & $0.035 \pm 0.001$ & $0.037 \pm 0.001$ & $0.030 \pm 0.000$ \\
    \midrule
    AIC & $x^\alpha$ & -193.22 & -279.36 & -438.15 & -432.59 & -254.52 & -172.20 & -1732.82 \\
    AIC & $e^{\alpha x}$ & -211.13 & -306.42 & -482.39 & -480.03 & -284.26 & -193.96 & -1887.22 \\
    AIC & $e^{\alpha x^2}$ & \textbf{-293.47} & \textbf{-436.46} & \textbf{-705.67} & \textbf{-746.36} & \textbf{-458.90} & \textbf{-326.59} & \textbf{-2286.23} \\
    AIC & $e^{\alpha x^3}$ & -278.35 & -410.57 & -663.73 & -621.05 & -354.01 & -228.19 & -2204.57 \\
    \midrule
    BIC & $x^\alpha$ & -191.81 & -277.55 & -435.86 & -430.30 & -252.74 & -170.80 & -1729.13 \\
    BIC & $e^{\alpha x}$ & -209.72 & -304.61 & -480.10 & -477.74 & -282.48 & -192.56 & -1883.53 \\
    BIC & $e^{\alpha x^2}$ & \textbf{-292.07} & \textbf{-434.65} & \textbf{-703.38} & \textbf{-744.07} & \textbf{-457.12} & \textbf{-325.18} & \textbf{-2282.54} \\
    BIC & $e^{\alpha x^3}$ & -276.95 & -408.77 & -661.44 & -618.76 & -352.23 & -226.79 & -2200.88 \\
    \midrule
    MSE & $x^\alpha$ & 0.01053 & 0.00830 & 0.00723 & 0.00732 & 0.00830 & 0.01037 & 0.00714 \\
    MSE & $e^{\alpha x}$ & 0.00731 & 0.00495 & 0.00384 & 0.00391 & 0.00495 & 0.00705 & 0.00374 \\
    MSE & $e^{\alpha x^2}$ & \textbf{0.00378} & \textbf{0.00130} & \textbf{0.00017} & \textbf{0.00021} & \textbf{0.00128} & \textbf{0.00333} & \textbf{0.00006} \\
    MSE & $e^{\alpha x^3}$ & 0.00413 & 0.00168 & 0.00057 & 0.00062 & 0.00165 & 0.00359 & 0.00047 \\
    \midrule
    \bottomrule
  \end{tabular}
  }
  \caption{Table of statistical fits for the expected number of energy evaluations in classical simulated annealing for the random dataset in 3D. In bold, we indicate the model preferred by a particular model selection criterion {(see Section \ref{section:methods})}.}
  \label{table:sa_random_3d}
\end{sidewaystable}

\end{document}